\documentclass[10pt]{iopart}
\usepackage{latexsym}
\usepackage{amssymb}
\usepackage{amsfonts}
\usepackage{theorem}
\usepackage{graphicx}
\usepackage{fullpage}
\usepackage[latin1]{inputenc}
\usepackage[T1]{fontenc}
\usepackage[english]{babel}

\def\be{\begin{equation}}
\def\ee{\end{equation}}
\def\bea{\begin{eqnarray}}
\def\eea{\end{eqnarray}}

\def\nn{\nonumber}

\newcommand{\R}{\mathbb{R}}

\newcommand{\E}{\mathbb{E}}

\newcommand{\Z}{\mathbb{Z}}

\newcommand{\bm}{\mbox{\boldmath $m$}}
\newcommand{\bxi}{\mbox{\boldmath $\xi$}}

\newcommand{\Nu}{\mathcal{N}}
\newcommand{\Hi}{\mathcal{H}}

\newcommand{\order}{O}

\newcommand{\bb}{\mbox{\boldmath $b$}}
\newcommand{\bz}{\mbox{\boldmath $z$}}
\newcommand{\bsigma}{\mbox{\boldmath $\sigma$}}
\newcommand{\blambda}{\mbox{\boldmath $\lambda$}}

\newcommand{\vsp}{\vspace*{3mm}}
\newcommand{\room}{\rule[-0.2cm]{0cm}{0.6cm}}

\def\a{\alpha}
\def\s{\sigma}

\def\b{\beta}
\def\g{\gamma}

\def\d{\delta}
\def\l{\lambda}

\begin{document}

\title{Immune networks: multi-tasking capabilities at medium load}

\author{E Agliari$^{1,2}$, A Annibale$^{3,4}$, A Barra$^{5}$, ACC Coolen$^{4,6}$, and D Tantari$^7$}

\address{$^1$ Dipartimento di Fisica, Universit\`{a} degli Studi di Parma, Viale GP Usberti 7/A,  43124 Parma, Italy}
\address{$^2$ INFN, Gruppo Collegato di Parma, Viale Parco Area delle Scienze 7/A,
 43100 Parma, Italy}
\address{$^3$ Department of Mathematics, King's College London,  The Strand, London WC2R 2LS, UK}
\address{$^4$ Institute for Mathematical and Molecular Biomedicine, King's College London, Hodgkin Building, London SE1 1UL, UK}
\address{$^5$ Dipartimento di Fisica, Sapienza Universit\`{a} di Roma, P.le Aldo Moro 2, 00185 Roma, Italy}
\address{$^6$ London Institute for Mathematical Sciences, 35a South St, Mayfair, London W1K 2XF, UK}
\address{$^7$ Dipartimento di Matematica, Sapienza Universit\`{a}  di Roma, P.le Aldo Moro 2, 00185 Roma  Italy}

\begin{abstract}
Associative network models featuring multi-tasking properties have been introduced recently and studied in the low load regime, where the number $P$ of simultaneously retrievable patterns scales with the number $N$ of nodes as $P\sim \log N$. In addition to their relevance in artificial intelligence, these models are increasingly important in immunology, where stored patterns represent strategies to fight pathogens and nodes represent lymphocyte clones.  They allow us to understand the crucial ability of the immune system to respond simultaneously to multiple distinct antigen invasions. Here we develop further  the statistical mechanical analysis of such systems,
by studying the medium load regime, $P \sim N^{\d}$ with $\d \in (0,1]$.
We derive three main results. First, we reveal the nontrivial architecture of these networks: they exhibit a high degree of modularity and clustering, which is linked to their retrieval abilities. Second, by solving the model we demonstrate for $\delta<1$ the existence of large  regions in the phase diagram where the network can retrieve all stored patterns simultaneously.  Finally, in the high load regime $\delta=1$ we find that the system behaves as a spin glass, suggesting that finite-connectivity frameworks are required to achieve effective
retrieval.
\end{abstract}


\ead{agliari@fis.unipr.it,alessia.annibale@kcl.ac.uk,adriano.barra@roma1.infn.it,\\ton.coolen@kcl.ac.uk,tantari@mat.uniroma1.it}

\section{Introduction}

After a pioneering paper \cite{parisi} followed by a long period of dormancy,
recent years have witnessed a surge of  interest in statistical mechanical models of the immune system
\cite{BA1,PRE,JTB1,bialek,kosmir1,kosmir2,chakra,PREdeutch2,ton6}. This description complements the more standard approaches, which tend to be phrased in the language of dynamical systems \cite{perelson,perelson2,din1,din2}.  To make further progress, however, it has become clear that  we need new quantitative tools, able to handle the complexities which surfaced in e.g. \cite{alps2_lett,alps2_lungo}.
 This is the motivation for the present study.

There is an intriguing and fruitful analogy (from a modelling perspective) between neural networks, which have been modelled in statistical mechanics quite extensively, and immune networks. Let us highlight the similarities and differences.
In neural networks the nodes represent neurons, which interact with each other directly through Hebbian synaptic couplings.  In (adaptive) immune systems, effector branches (B-clones) and coordinator branches (helper and suppressor T-clones), interact
via  signaling proteins called cytokines. The latter can represent both eliciting and suppressive signals.
Neural and immune systems are both able to learn (e.g. how to fight new antigens), memorize (e.g. previously encountered antigens) and  `think'  (e.g. select the best strategy to cope with pathogens).
However, neural networks are designed for serial processing:  neurons perform collectively to retrieve a {\em single} pattern at a time. This is not acceptable in the immune context. Multiple antigens will normally be present at the same time, which requires the simultaneous recall of multiple patterns (i.e. of multiple defense strategies). Moreover, the architectures of neural and immune networks are very different. A model with fully connected topology, mathematically convenient but without a basis in biological reality, is tolerable for neural networks where each neuron is known to have a huge number of connections with others. In contrast, in immune networks, where interactions among lymphocytes are much more specific, the underlying topology must be carefully modelled and is expected to play a crucial operational role.
From a theoretical physics perspective, a network of interacting B- and T-cells  resembles a bipartite spin glass.
It was recently shown that such bipartite spin glasses exhibit retrieval features which are deeply related to their structures \cite{alps2_lett,alps2_lungo}, and this can be summarized as follows:
\begin{itemize}

\item There exists a structural equivalence between Hopfield neural networks and bipartite spin glasses.  In particular, the two systems share the same partition function, and hence the same thermodynamics \cite{BGG,hotelNN}.

\item One can either dilute directly a Hopfield network, or its underlying bipartite spin glass. The former does not affect pattern retrieval qualitatively \cite{sompo1,ton0,ton1,ton3,sompo2,ton4}, whereas the latter causes a switch from serial to  parallel processing \cite{alps2_lett,alps2_lungo} (i.e. to simultaneous pattern recall).

\item Simultaneous pattern recall is essential in the context of immunology, since it implies the ability to respond to multiple antigens simultaneously. The analysis of such systems requires a combination of techniques from statistical mechanics and graph theory.

\end{itemize}
The last point is the focus of the present paper, which is organized as follows. In Section 2 we describe a minimal biological scenario for the immune system, based on the analogy with neural networks. We define our model and its scaling regimes, and prepare the stage for calculations. Section 3 gives a comprehensive analysis  of the topological properties of the network in the extremely diluted regime, which is the scaling regime assumed throughout our paper. Section 4 is dedicated to the statistical mechanical analysis of the system in the medium load regime, focusing on  simultaneous pattern recall of the network. Section 5 deals with the high load regime. Here the network is found to behave as a spin glass,
suggesting that a higher degree of dilution should be implemented -- in remarkable agreement with immunological findings \cite{janaway,abbas} --
and this will be the focus of future research. The final section gives a summary of our main conclusions.

\section{Statistical mechanical modelling of the adaptive immune system}

\subsection{The underlying biology}

All mammals have an innate (broad range) immunity, managed by macrophages, neutrophils, etc., and an adaptive immune response. The latter is highly specific for particular targets, handled by lymphocytes, and the focus of this paper.
To be concise, the following introduction to the adaptive immune system has already been filtered by a theoretical physics perspective, and immunological observables are expressed in`physical' language. We refer to the excellent books \cite{janaway,abbas} for comprehensive  reviews of the immune system, and to a selection of papers \cite{BA1,PRE,JTB1,alps2_lett,alps2_lungo,BA2} for explanations of the link between `physical' models and biological reality.
Our prime interest is in B-cells and in T-cells; in particular, among T-cells, in the subgroups of so-called `helpers' and `suppressors'.
B-cells produce antibodies and present them on  their external surface in such a way that they are able to recognize and bind pathogenic peptides. All B-cells that produce the same antibody belong to the same clone, and the ensemble of all the different clones forms the immune  repertoire. This repertoire is  of size $\mathcal{O}(10^8-10^9)$ clones in humans. The size of a clone, i.e. the number of identical B-cells, may vary strongly. A clone at rest may contain some $\mathcal{O}(10^3-10^4)$ cells, but when it undergoes clonal expansion its size may increase by several orders of magnitude, to up to $\mathcal{O}(10^6-10^7)$. Beyond this size the state of the immune system would be pathological, and is referred to as lymphocytosis.

When an antigen enters the body,  several antibodies (i.e. several B-cells belonging to different clones) may be able to bind to it, making it chemically inert and biologically inoffensive. In this case, conditional on authorization by T-helpers (mediated via cytokines), the binding clones undergo clonal expansion. This means that their cells start duplicating, and  releasing high quantities of soluble antibodies to inhibit the enemy. After the antigen has been deleted, B-cells are instructed by T-suppressors, again via cytokines, to stop producing antibodies and undergo apoptosis. In this way the clones reduce their sizes, and order is restored.
Thus, two signals are required for B-cells to start clonal expansion: the first signal is binding to antigen, the second is a `consensus' signal, in the form of an eliciting cytokine secreted by T-helpers. This latter mechanism prevents abnormal reactions, such as autoimmune manifestations\footnote{Through a phenomenon called `cross-linking', a B-cell can also have the ability to bind a self-peptide, and may accidentally start duplication and antibody release, which is a dangerous unwanted outcome.}.

T-helpers and T-suppressors are lymphocytes that work `behind the scenes', regulating the immune response by coordinating the work of effector branches, which in this paper are the B-cells.
To accomplish this, they are able to secrete both stimulatory and suppressive chemical signals, the cytokines \cite{chitochine,chitochinebook}. If within a given (small)  time interval a B-clone recognizes an antigen and detects an eliciting cytokine from a T-cell, it will become activated
 and start duplicating and secreting antibodies.  This scenario is the so-called `two-signal model' \cite{goodnow1,goodnow2,goodnow3,anergy3}. Conversely, when the antigen is absent and/or the cytokine signalling is suppressive, the  B-cells tuned to this antigen start the apoptosis programme, and their immuno-surveillance is turned down to a rest state. For simplicity, we will from now on with the term `helper' indicate any helper or suppressor T-cell.
The focus of this study is to understand, from a statistical mechanics perspective, the ability of helpers and suppressors to coordinate and manage {\em simultaneously} a huge ensemble of B-clones (possibly {\em all}).

\subsection{A minimal model} \label{sec:formal}

\begin{figure}[t]
\unitlength=0.2mm
 \begin{center}{
\begin{picture}(500,300)
\put(0,0){\includegraphics[width=440\unitlength,height=270\unitlength]{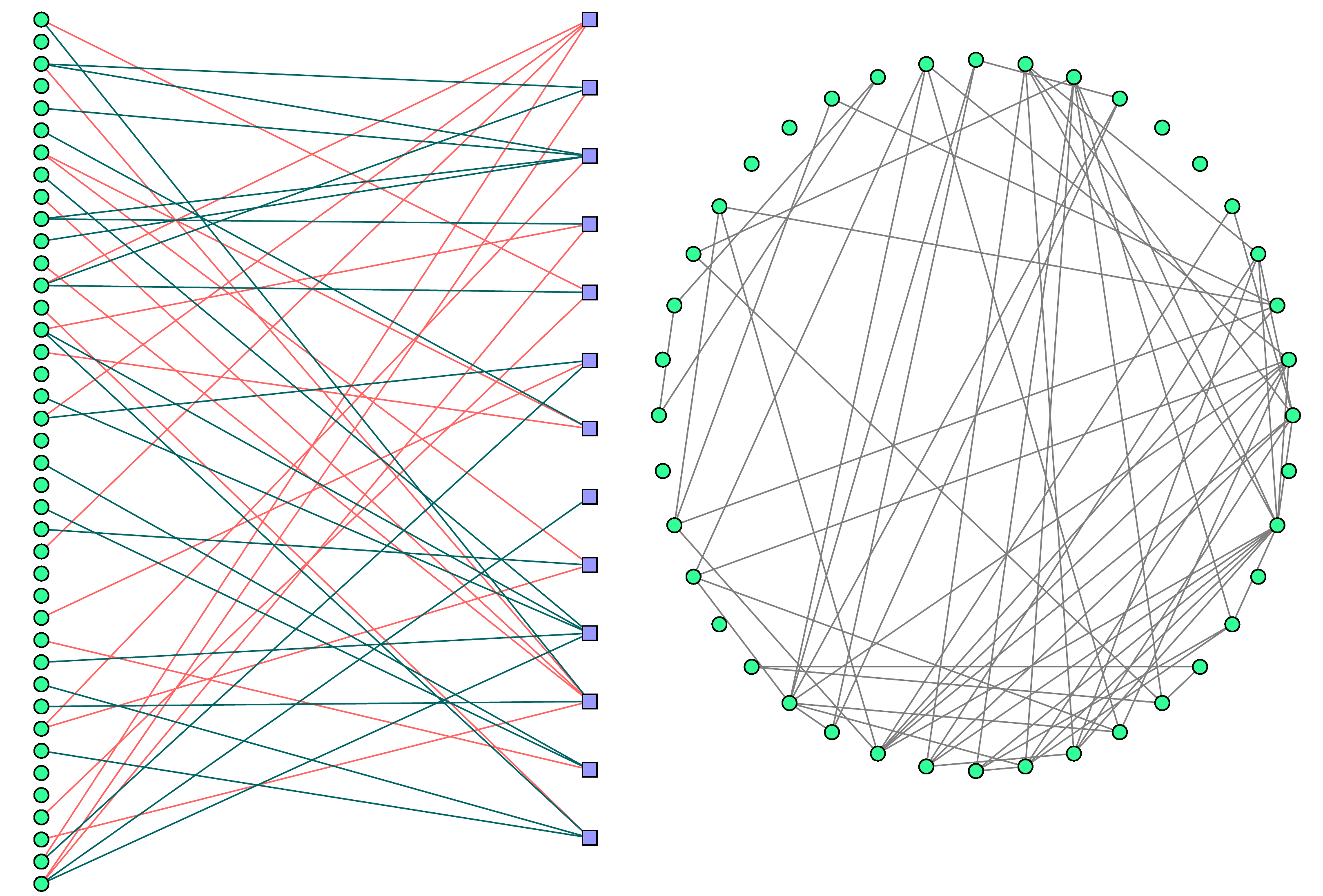}}
\put(-10,290){\em T-cells} \put(175,290){\em B-cells}
\put(455,140){\em T-cells only}
\end{picture}
}
\end{center}
\caption{\label{fig:system} Left: schematic representation of the bipartite spin-glass which models the interaction between B- and T-cells through cytokines. The latter are drawn as colored links,  with red representing stimulatory cytokines (positive couplings) and
 black representing inhibiting ones (negative couplings). Note that the network is diluted. Right: the equivalent associative multitasking network consisting of T-cells only, obtained by integrating out the B-cells. This network is also diluted, with links given by the Hebbian prescription.}
\label{fig:map}
\end{figure}
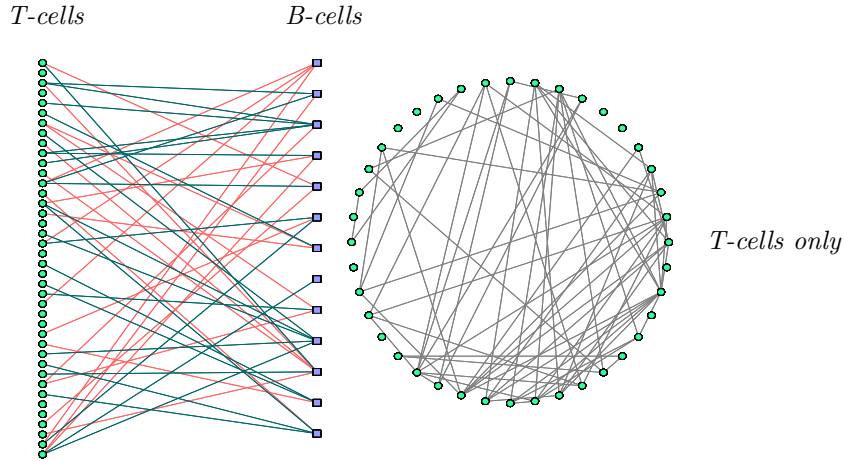

We consider an immune repertoire of $N_B$ different clones, labelled by $\mu \in \{1,...,N_B\}$. The size of  clone $\mu$ is $b_{\mu}$.
In the absence of interactions with helpers, we take the clone sizes to be Gaussian distributed; without loss of generality we may take the mean to be zero  and unit width, so $b_{\mu} \sim \mathcal{N}(0,1)$. A value $b_{\mu}\gg 0$ now implies that clone $\mu$ has expanded (relative to the typical clonal size), while $b_{\mu}\ll 0$ implies  inhibition.
The Gaussian clone size distribution is supported both by experiments and by theoretical arguments \cite{JTB1}.
Similarly, we imagine having $N_T$ helper clones, labelled by $i \in \{1,...,N_T\}$.
The state of helper clone $i$ is denoted by $\s_i$. For simplicity, helpers are assumed to be in only two possible states: secreting cytokines ($\s_i=+1$) or quiescent ($\s_i=-1$).  Clone sizes $b_\mu$ and the helper states $\s_i$ are dynamical variables.
We will abbreviate $\bsigma=(\s_1,\ldots,\s_{N_T})\in\{-1,1\}^{N_T}$, and $\bb=(b_1,\ldots,b_{N_B})\in\R^{N_B}$.

The interaction between the helpers and B-clones is implemented by cytokines. These are taken to be frozen (quenched) discrete  variables.
The effect of a  cytokine secreted by helper $i$ and detected by clone $\mu$ can be nonexistent ($\xi_i^\mu=0$), excitatory ($\xi_i^{\mu} = 1$), or inhibitory
 ($\xi_i^{\mu} = -1$). To achieve a Hamiltonian formulation of the system, and  thereby enable equilibrium analysis, we have to impose symmetry of the cytokine interactions. So, in addition to the B-clones being influenced by cytokine signals from helpers, the helpers will similarly feel a signal from the B-clones.  This symmetry assumption can be viewed as a necessary first step, to be relaxed in future investigations, similar in spirit to the early formulation of symmetric spin-glass models for neural networks \cite{DGZ,ton5}.  We are then led to  a Hamiltonian $\hat{\mathcal{H}}(\bb,\bsigma|\xi)$ for the combined system of the following form (modulo trivial multiplicative factors):
\be
\hat{\mathcal{H}}(\bb,\bsigma|\xi) = -\frac{1}{\sqrt{N_T}}\sum_{i=1}^{N_T}\sum_{\mu=1}^{N_B}\xi_i^{\mu}\s_i b_{\mu} +\frac{1}{2\sqrt{\beta}}\sum_{\mu=1}^{N_B}b_\mu^2,
\ee
In the language of disordered systems, this is a bipartite spin-glass. We can integrate out the variables $b_\mu$, and map our system to a model with helper-helper interactions only.
The partition function $Z_{N_T}(\beta,\xi)$, at inverse clone size noise level $\sqrt{\beta}$ (which is the level consistent with our assumption
$b_{\mu} \sim \mathcal{N}(0,1)$) follows straightforwardly,  and reveals the mathematical equivalence with an associative attractor network:
\begin{eqnarray}
Z_{N_T}(\beta,\xi) &=& \sum_{\bsigma}\int\!db_1\ldots db_{N_B}\exp[-\sqrt{\beta}~\hat{\mathcal{H}}(\bb,\bsigma|\xi) ]
\nonumber\\
&=& \sum_{\bsigma}\exp [-\beta \mathcal{H}(\bsigma|\xi)],
\label{equivalence}
\end{eqnarray}
in which, modulo an irrelevant additive constant,
\begin{eqnarray}
\mathcal{H}(\bsigma|\xi)
&=&-\frac{1}{2}\sum_{ij=1}^{N_T}\s_i J_{ij}\s_j,~~~~~~J_{ij}=\frac{1}{N_T}\sum_{\mu=1}^{N_B}\xi_i^{\mu}\xi_j^{\mu}
\label{eq:hopfield}
\end{eqnarray}
Thus, the system with Hamiltonian $\hat{\mathcal{H}}(\bb,\bsigma|\xi)$, where helpers and B-clones interact through cytokines, is thermodynamically equivalent to a Hopfield-type associative network represented by $\mathcal{H}(\bsigma|\xi)$, in which helpers mutually interact through an effective Hebbian coupling. See Figure \ref{fig:map}. Learning a pattern in this model then means adding a new B-clone with an associated string of new cytokine variables.

If there are no zero values for the $\{\xi_i^\mu\}$, the system characterized by (\ref{eq:hopfield}) is well known in artificial intelligence research.  It is able to retrieve each of the $N_B$ `patterns' $(\xi_1^\mu,\ldots,\xi_N^\mu)$, provided these patterns are sufficiently uncorrelated, and both the ratio $\alpha=N_B/N_T$ and the noise level $1/\beta$ are sufficiently small  \cite{JTB1,ton0,amit,ton2}.
Retrieval quality can be quantified by introducing $N_B$ suitable order parameters, viz. $m_{\mu}(\bsigma)= N_T^{-1}\sum_i \xi_i^{\mu}\s_i$, in terms of which  the new Hamiltonian (\ref{eq:hopfield})  can be written as
\begin{equation} \label{eq:intemedio}
\mathcal{H}(\bsigma|\xi)= -\frac{N_T}{2}\sum_{\mu=1}^{N_T} m_{\mu}^2(\bsigma).
\end{equation}
If $\alpha$ is sufficiently small, the minimum energy configurations of the system are those where  $m_{\mu}(\bsigma)=1$ for some $\mu$ (`pure states'),  which implies that $\bsigma=(\xi_1^\mu,\ldots,\xi_N^\mu)$ and  pattern $\mu$ is said to be retrieved perfectly.
But what does retrieval mean in our immunological context?  If $m_\mu(\bsigma)=1$, all the helpers are `aligned' with their coupled cytokines: those $i$ that inhibit clone $\mu$ (i.e. secrete $\xi_{i}^\mu=-1$)  will be quiescent ($\s_{i}=-1$), and
those $i$ that excite clone $\mu$ (i.e. secrete $\xi_{i}^\mu=1$)  will be active ($\s_{i}=1$) and release the eliciting cytokine. As a result the B-clone $\mu$ receives the strongest possible positive signal (i.e. the random environment becomes a `staggered magnetic field'), hence it is forced to expand. Thus the arrangement of helper cells leading to the retrieval of pattern $\mu$ corresponds to clone-specific excitatory signalling upon the B-clone $\mu$.

However, if all $\xi_i^\mu\in\{-1,1\}$ so the bipartite network  is fully connected, it can expand only one B-clone at a time. This would be a disaster for the immune system. We need  the dilution in the bipartite B-H network that is caused by having also  $\xi_i^\mu=0$ (i.e. no  signalling between helper $i$ and clone $\mu$), to  enable multiple clonal expansions.
The associative network (\ref{eq:hopfield}) now involves patterns with blank entries, and `pure states'  no longer work as low energy configurations. Retrieving a pattern no longer employs all spins $\sigma_i$, and those corresponding to null entries  can be used to recall other patterns. This is energetically favorable since the energy is quadratic  in the magnetizations $m_{\mu}(\bsigma)$.
Conceptually, this is only a reshaping of the network's recall tasks: no  theoretical bound for information content is violated, and  global retrieval is still performed through $N_B$ bits. However, the perspective is shifted: the system no longer requires a sharp resolution in information exchange between a helper clone and a B-clone\footnote{In fact, the high-resolution analysis is performed in the antigenic recognition on the B-cell surface, which is based on a sharp key-and-lock mechanism \cite{BA1}.}. It suffices that a B-clone receives an attack signal, which could be encoded  even by a single bit. In a diluted bipartite B-H system the associative capabilities of the helper network are distributed, in order  to simultaneously manage the whole ensemble of B-cells.
The analysis of these immunologically most relevant pattern-diluted versions of associative networks is still at an early stage.  So far only the low storage case $N_B \sim \log N_T$ has been solved \cite{alps2_lett,alps2_lungo}. In this paper we analyse
the extreme dilution regime for the B-H system,  i.e. $N_B \sim N_T^{\delta}$ with $0<\delta\leq 1$.

\section{Topological properties of the emergent networks} \label{sec:topo}

\subsection{Definitions and simple characteristics}

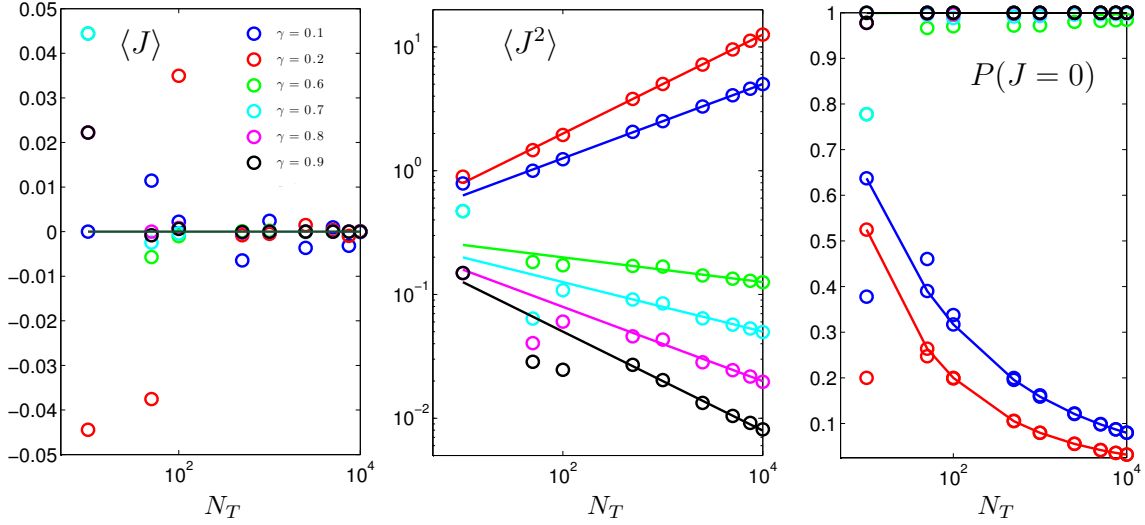
\begin{figure}[t]
\unitlength=0.19mm
\hspace*{-0mm}
\begin{picture}(800,350)
\put(0,0){\includegraphics[width=820\unitlength,height=370\unitlength]{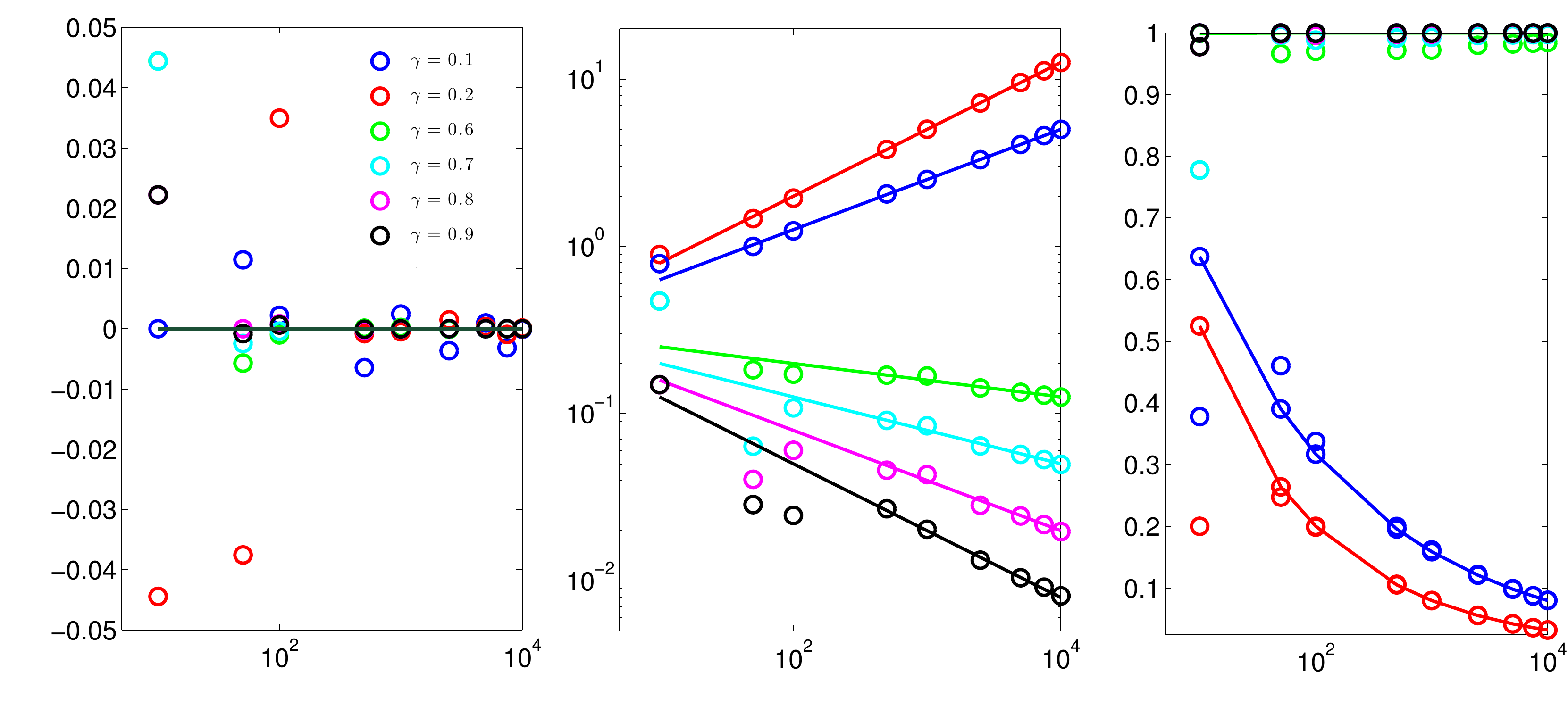}}
\put(164,2){$N_T$} \put(432,2){$N_T$} \put(705,2){$N_T$}
\put(104,325){\large $\langle J\rangle$} \put(372,325){\large $\langle J^2\rangle$} \put(700,302){\large $P(J=0)$}

\end{picture}
\vspace*{0mm}

\caption{\label{fig:check} Statistical properties of individual links in randomly generated instances of the graph $\mathcal{G}$ at different sizes $N_T$, with $N_B=\alpha N_T^{\d}$.  We measured the mean coupling $\langle J \rangle$ (left), the mean squared coupling $\langle J^2 \rangle$ (middle) and the probability $P(J=0)$ of a zero link (right), for  different values of $\gamma$. The parameters  $\d=1$ and $\alpha=0.5$ are kept fixed. Solid lines:  predictions given in  (\ref{eq:coupling_estimate1}) and (\ref{eq:P0_estimate}). Markers: simulation data.
}
\end{figure}

We start with the definition of the bi-partite graph,  which contains two sets of nodes (or vertices): the set
$V_B$ representing B-cells (labelled by $\mu$) and the set $V_T$ representing T-cells (labelled by $i$), of cardinality $N_B$ and $N_T$, respectively. Nodes belonging to different sets can be pairwise connected via links, which are identically and independently drawn with probability $p$, in such a way that a random bipartite network $\mathcal{B}$ is built. We associate with each link a weight, which can be either $+1$ or $-1$; these weights are quenched and drawn randomly from a uniform distribution.
As a result, the state of each link connecting the $\mu$-th B-clone and the $i$-th T-clone can be denoted by a random variable $\xi_i^{\mu}$, distributed independently according to
\begin{eqnarray} \label{eq:xi}
P(\xi^{\mu}_i)&=&\frac{p}{2}(\delta_{\xi_i^\mu,1}+ \delta_{\xi_i^\mu,-1})+(1\!-\!p)\delta_{\xi_i^\mu,0}
\end{eqnarray}
We choose $p=c/N_T^\gamma$, with $\gamma \in [0, \infty)$ subject to $p\leq 1$, and $c=\order(N_T^0)$.
Upon tuning $\gamma$, $\mathcal{B}$ displays different topologies, ranging from fully connected (all $N_T \times N_B$ possible links are present, for $\gamma\to 0$) to fully disconnected (for $\gamma\to\infty$).
We have shown in the previous section how a process on this bipartite graph can be mapped to  a thermodynamically equivalent process  on a new graph, built  only of the $N_T$ nodes in $V_T$, occupied by spins $\sigma_i$ that  interact pairwise through a coupling matrix with (correlated) entries
\be
J_{ij} = \sum_{\mu = 1}^{N_B} \xi_i^{\mu} \xi_j^{\mu}.
\label{eq:Jij}
\ee
The structure of the marginalized system is represented by a weighted mono-partite graph $\mathcal{G}$, with weights (\ref{eq:Jij}), whose topology is controlled by $\gamma$. To illustrate  this, let us consider the weight distribution $P(J | N_B, N_T, \gamma, c)$, which can be interpreted as the probability distribution for the end-to-end distance of a one-dimensional random walk. This walk  has a waiting probability $p_w=1-p$, and probabilities of moving left ($p_l$) or right ($p_r$) equal to $p_l = p_r = p/2$, i.e.
\be
p_w = 1 - \left (c/N_T^{\gamma} \right )^2,~~~~~~
p_l = p_r = \frac{1}{2} \left ( c/N_T^{\gamma} \right )^2.
\ee
Therefore, we can write
\be \label{eq:RW}
P(J | N_B, N_T, \gamma, c) = {\sum_{S=0}^{L-J}}~{\large\hspace*{-1.5mm}^\prime} \frac{N_B!} {S! \left( \frac{N_B-S-J}{2} \right)! \left( \frac{N_B-S+J}{2} \right)!} ~ p_w^S \, p_r^{(N_B-S+J)/2} \, p_l^{(N_B-S-J)/2},
\ee
where the prime indicates that the  sum runs only over values of $S$ with the same parity as $N_B \!\pm \!J$. The result (\ref{eq:RW}) can easily be generalized to the case of biased weight distributions \cite{Amit-PRA1987}, which would correspond to non-isotropic random walks.
The first two moments of (\ref{eq:RW}) are, as confirmed numerically in Figure \ref{fig:check}:
\begin{eqnarray}\label{eq:coupling_estimate1}
\langle{J} \rangle = 0,~~~~~~
\langle{J^2} \rangle = \left ( c/N_T^{\gamma} \right )^2 N_B,
\label{eq:coupling_estimate2}
\end{eqnarray}
We now fix a scaling law for $N_B$, namely $N_B = \alpha N_T^{\delta}$, with $\alpha >0$. This includes the high-load regime for $\d=1$, as well as the medium-load regime for $\d \in (0,1)$. The low-storage regime $\d=0$ has already been treated elsewhere \cite{alps2_lett,alps2_lungo}.
We then find

\begin{figure}[t]
\unitlength=0.20mm
\hspace*{-10mm}
\begin{picture}(800,450)
\put(0,0){\includegraphics[width=880\unitlength,height=470\unitlength]{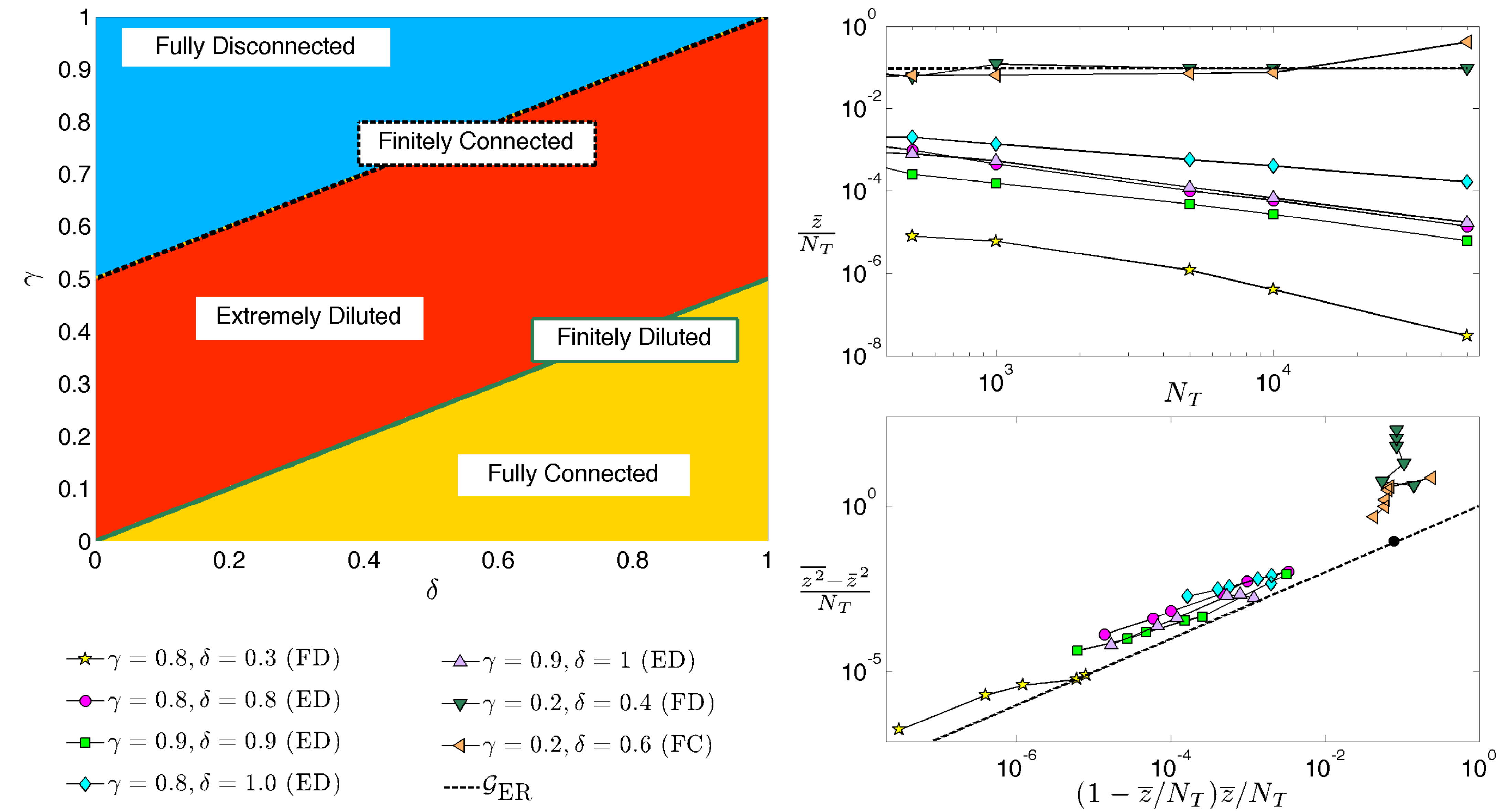}}

\end{picture}
\vspace*{1mm}

\caption{\label{fig:PhaseDiag} Left: qualitative phase diagram describing the different topological regimes of $\mathcal{G}$, according to the mean-field analysis in Sec.~\ref{sec:topo}. Right panels: finite-size scaling for the average degree $\bar{z}$ (upper) and fluctuations $\overline{z^2} - \bar{z}^2$ (lower), measured on realizations of $\mathcal{G}$ for different choices of parameters $\gamma$ and $\delta$ (see legend). The parameters $c=1$ and $\alpha=0.1$ are kept fixed,. The markers correspond to numerical data, and  the  lines connecting the markers are guides to the eye. The cases being compared include  fully disconnected (FD), extremely diluted (ED), finitely diluted (FD) and  fully connected (FC) regimes, and they are shown together with data  on Erd\"{o}s-R\`enyi graphs $\mathcal{G}_{\textrm{ER}}$ with link probability $q=1-e^{\alpha c^2}$ (dashed lines, see Eq.~\ref{eq:P0_estimate}). For ER graphs one expects $\bar{z}/N_T$ to be constant, and equal to $q$, while the normalized  fluctuations are expected to be $(1-\bar{z}/N_T)\bar{z}/N_T$. Different points pertain to different graph sizes, but with the same link probability $q$; they are found to overlap regardless of their size ($\bullet$). For $\mathcal{G}$, the behaviour of $\bar{z}$ is consistent with mean-field expectations, while connectivity fluctuations are underestimated by the mean-field approach.}
\vspace*{0mm}
\end{figure}

\be
\langle{J^2} \rangle =  \alpha c^2/N_T^{2\gamma-\d},
\ee
The probability of having $J \neq 0$ scales like $N_T^{\d - 2\gamma}$ for $2\gamma >\d$, while for $2\gamma < \d$ it approaches $1$ in the limit $N_T\to\infty$.
One can recover the same result via a simple approximation, which is valid in the case $p_r \ll 1$:
\begin{equation}
P(J=0 | N_B, N_T, \gamma, c) ~\approx ~p_w^{N_B} = \left[ 1 - \left ( c/N_T^{\gamma} \right )^2 \right]^{N_B},
\end{equation}
see also \ref{sec:topo2} for a more rigorous derivation of $P(J=0 | N_B, N_T, \gamma, c)$.
Given the assumed scaling of $N_B$, we get
$P(J \neq 0 | N_B(\alpha, \delta ,N_T), N_T, \gamma, c) \approx 1 - e^{- \alpha c^2 {N_T}^{\delta - 2 \gamma}}$, which translates into
\begin{equation} \label{eq:P0_estimate}
P(J \neq 0 | N_B(\alpha, \d,N_T), N_T, \gamma, c) \approx
\left\{
\begin{array}{cl}
\alpha c^2 {N_T}^{\d - 2 \gamma} & {\rm if}~~2 \gamma > \d \\
1 - e^{-\alpha c^2} & {\rm if}~~2 \gamma= \d \\
1 &  {\rm if}~~2 \gamma < \d \\
\end{array}
\right.,
\end{equation}
This quantity can be interpreted as the average link probability in $\mathcal{G}$. The average degree $\bar{z}$ over the whole set of nodes\footnote{The degree or coordination number $z_i$ of node $i$ is the number of its nearest-neighbors, i.e. the number of links stemming from the node itself. Thus, the average degree $\bar{z}= \sum_{i \in V_T} z_i/N_T$ measures the density of links present in the graph.} can then be written as
\be \label{eq:z_estimate}
\bar{z} = N_T P(J \neq 0).
\ee
Thus, if we adopt a mean-field approach based only on the estimates (\ref{eq:P0_estimate},\ref{eq:z_estimate}), we find that $\mathcal{G}$ can display the following topologies, expressed in terms of the average degree $\overline{z}$ of $\mathcal{G}$ (the average number of links per node):\vspace*{9mm}

{
\begin{tabular}{l|ll}
\hline
\room
& $0<\gamma<1$   & $\gamma=1$ \\
\hline
\room $\delta<2\gamma-1$ & fully disconnected, $\overline{z}\to 0$         &  fully disconnected, $\overline{z}\to 0$  \\
\room $\delta=2\gamma-1$ & finitely connected, $\overline{z}=\order(1)$  & finitely connected, $\overline{z}=\order(1)$  \\
\room $2\gamma-1<\delta<2\gamma$ & extremely diluted, $\overline{z}\to\infty$ but $\overline{z}/N_T\to 0$ & ~~~~~~------\\
\room $\delta=2\gamma$ & finitely diluted, $\overline{z}=\order(N_T)$      & ~~~~~~------\\
\room $\delta>2\gamma$ & fully connected, $\overline{z}=N_T$  &  ~~~~~~------\\
\hline
\end{tabular}
}
\vspace*{9mm}

\noindent
The missing entries in the table correspond to forbidden values $\delta\notin(0,1]$.
The various cases are also summarized in the left panel of Fig.~\ref{fig:PhaseDiag}. This picture is confirmed by numerical simulations. The right panels of Fig.~\ref{fig:PhaseDiag} give a finite size scaling analysis for the average degree $\bar{z}$ and its fluctuations $\overline{z^2} - \bar{z}^2$, measured in realizations of $\mathcal{G}$ for several choices of parameters. We also show corresponding data for  Erd\"{o}s-R\'{e}nyi graphs, where all links are independently drawn with probability $q$,  for comparison (here $\bar{z}=q N_T$ and $\bar{z^2} - \bar{z}^2=N_T q(1-q)$). We find that (i) in the fully disconnected regime of the phase diagram, $\bar{z}$ decays to zero exponentially as a function of $N_T$, (ii)
in the extremely diluted regime, $\bar{z}$ scales with $N_T$ according to a power law, (iii)  in the finitely diluted regime $\bar{z}$ is proportional to $N_T$, and (iv)  in the fully connected regime $\bar{z}$ saturates to $N_T$. There is thus full agreement with the predictions of the mean-field approach. The  fluctuations are slightly larger that those of a  purely randomly drawn network, which  suggests that $\mathcal{G}$ exhibits a certain degree of inhomogeneity. This will be investigated next.

It is important to stress that, as the system parameters $\gamma$ and $\delta$ are tuned, the connectivity of the resulting network $\mathcal{G}$ can vary \emph{extensively} and therefore, in order for the Hamiltonian (\ref{eq:intemedio}) to scale linearly with the system size $N_T$, the prefactor $1/N_T$ of the Hopfield model embedded in complete graphs, is not generally appropriate. One should normalize $\mathcal{H}(\bsigma|\xi)$ according to the expected connectivity of the graph.

\begin{figure}[t]
\unitlength=0.29mm
\hspace*{-3.5mm}
\begin{picture}(600,210)
\put(0,0){\includegraphics[width=280\unitlength,height=220\unitlength]{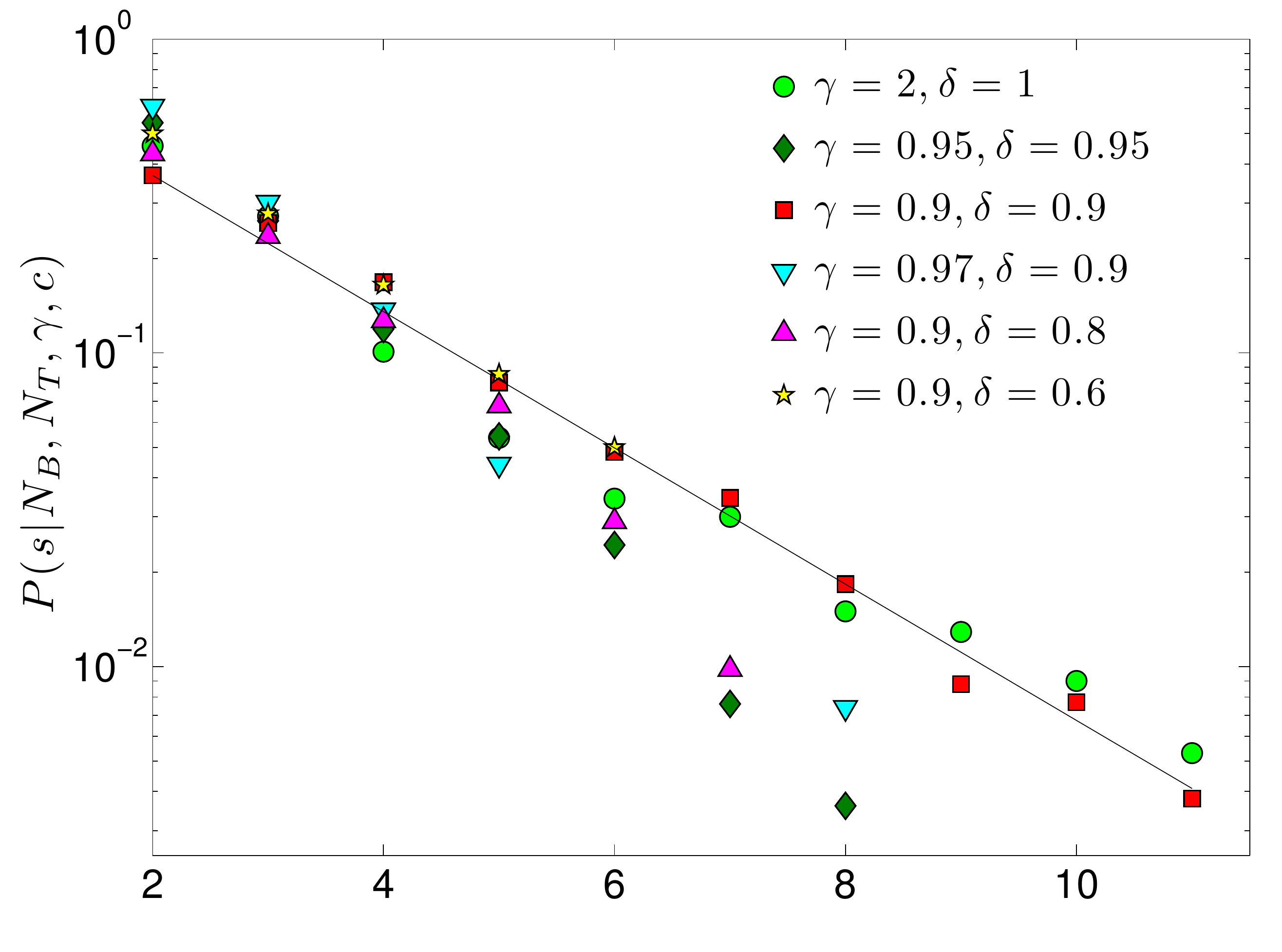}}
\put(300,6){\includegraphics[width=280\unitlength,height=217\unitlength]{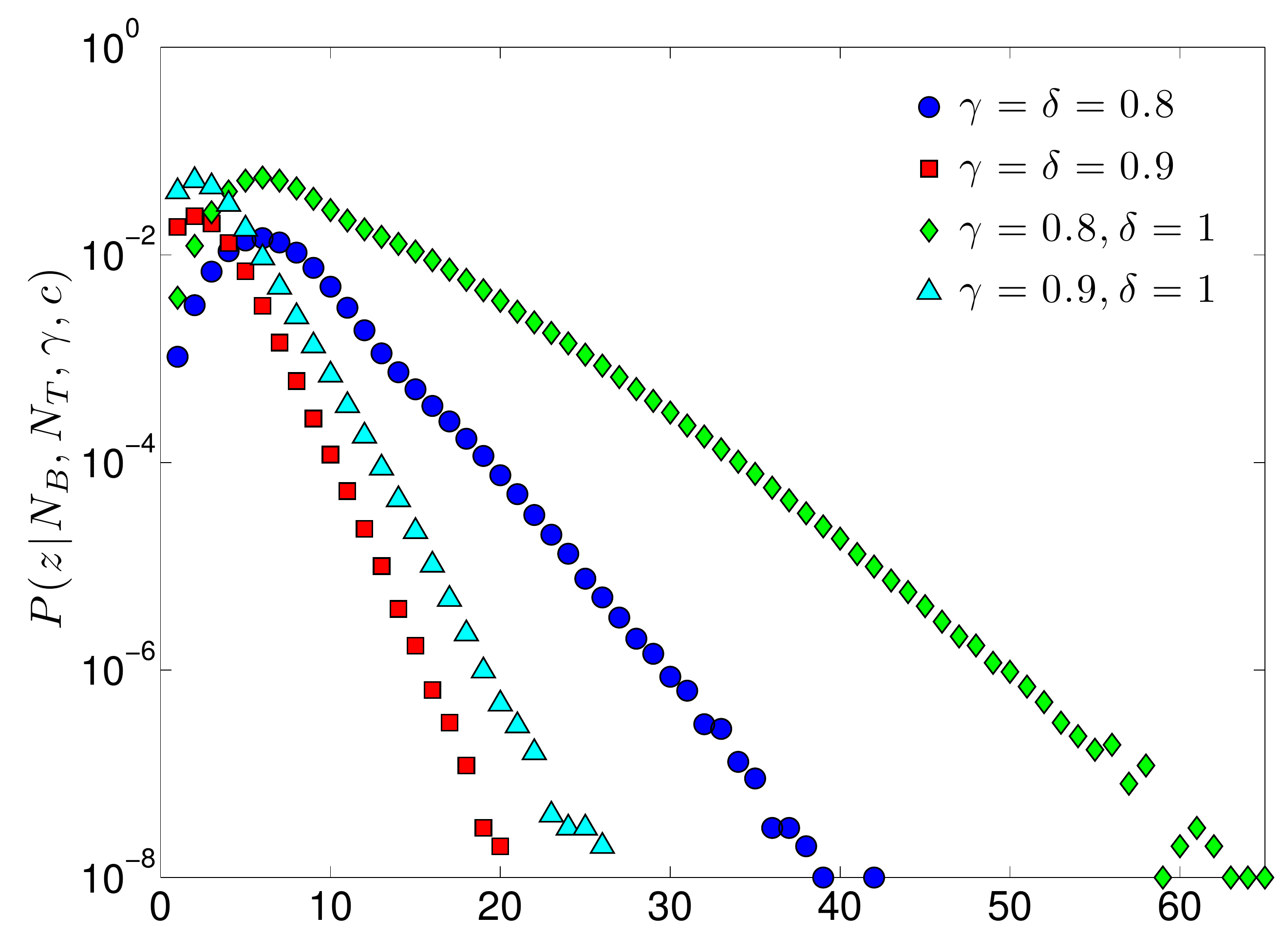}}
\put(150,-2){\large $s$} \put(450,-2){\large $z$}
\end{picture}

\caption{\label{fig:comp} Left: distribution of component sizes of the graph $\mathcal{G}$, for $\delta=c=1$, $\alpha =0.1$ and different values of $\gamma$. Symbols represent simulation data; the solid line represents the analytical estimate for the underpercolated case ($\square$). Right: degree distributions obtained for different values of $\delta$ and $\gamma$, with $\alpha = 0.1$ and $c=1$ kept fixed. The semi-logarithmic scale highlights the exponential decay at large values of $z$.}

\end{figure}

\subsection{Component size distribution} \label{sec:component}

As $\gamma$ is increased, both $\mathcal{B}$ and $\mathcal{G}$ become more and more diluted, eventually under-percolating.
The topology analysis can be carried out more rigorously for the (bi-partite) graph $\mathcal{B}$, since its link probability $p=c/N_T^{\gamma}$ is constant and identical for all $(i,\mu)$.
We can apply the generating function formalism developed in \cite{newman,allard} to show that the size of the giant component ($\subseteq V_T, V_B$) diverges when
\be
p^2 = 1/N_B N_T.
\ee
Hence, upon setting $N_B= \alpha N_T^{\d}$, the percolation threshold for the bipartite graph $\mathcal{B}$ is defined by
\be
N_T^{2 \gamma -1-\d} = c^2 \alpha = \order(N_T^0) ~~~\Rightarrow~~~\gamma=(\d+1)/2,
\ee
which is
consistent with the results of Section \ref{sec:topo}; we refer to \ref{genfunc} for full details.
Below the percolation threshold the generating function formalism also allows us to get the distribution $P_{\mathcal{B}}(s|N_T,N_B,c,\gamma)$ for the size $s$ of the small components occurring in $\mathcal{B}$. A (connected) component of an undirected graph is an isolated subgraph in which any two vertices are connected to each other by paths; the size of the component is simply the number of nodes belonging to the component itself. We prove in \ref{genfunc} that just below the percolation threshold, $P_{\mathcal{B}}(s|N_B,N_T,c,\gamma)$ scales exponentially with $s$. One finds that this is true also for the distribution $P_{\mathcal{G}}(s|N_B,N_T,c,\gamma)$ of graph ${\mathcal{G}}$ (see Figure \ref{fig:comp}, left panel).

Interestingly, the small components of $\mathcal{G}$ that emerge around and below the percolation threshold play a central role in the network's retrieval performance. To see this, one may consider the extreme case where the bi-partite graph $\mathcal{B}$ consists of trimers only. Here each node $\mu \in V_B$ is connected to two nodes $i_1, i_2 \in V_T$, that is $|\xi_{i_1}^{\mu}| = |\xi_{i_2}^{\mu}| = 1$ and
$\xi_{i_1}^{\nu} = \xi_{i_2}^{\nu}=0, \forall \nu \neq \mu$. The associated graph $\mathcal{G}$ is then made up of dimers $(i_1,i_2)$ only, and $J_{ij}
\in \{-1,0,1\}$ for all $(i,j)$. The energetically favorable helper cell configuration $\bsigma$ is now the one where $\sum_{i} \xi_i^{\mu} \sigma_i = \pm 2$, for any $\mu$. This implies that retrieval of all patterns is accomplished (under proper normalization). In the opposite extreme case, $\mathcal{B}$ is fully connected, and the helper cell system becomes a Hopfield network where parallel retrieval is not realized.
In general, around and below the percolation threshold, the matrix $\xi$ turns out to be partitioned, which implies that also the coupling matrix $\mathbf{J}$ is partitioned, and each block of $\mathbf{J}$ corresponds to a separate component of the overall graph $\mathcal{G}$. For instance, looking at the bipartite graph $\mathcal{B}$, a star-like module  with node $\mu \in V_B$ at its center and  the nodes $i_1,i_2,...,i_n \in V_T$ as leaves\footnote{The opposite case of a star-like module with the center belonging to $V_T$ is unlikely, given that $N_T>N_B$.} can occur when the leaves share a unique non-null $\mu$-th entry in their patterns, that is $|\xi_{i_1}^{\mu}| = |\xi_{i_2}^{\mu}| = ... = |\xi_{i_n}^{\mu}| =1$. For the graph $\mathcal{G}$ this module corresponds to a complete sub-graph $K_n$ of $n \leq N_T$ nodes. In this case the retrieval of pattern $\mu$ is trivially achieved.
In fact, a complete sub-graph $K_n$ in $\mathcal{G}$ can originate from more general arrangements in $\mathcal{B}$: each leaf $i$ can display several other null entries beyond $\mu$, but these are not shared, that is $\xi_i^{\nu} \xi_j^{\nu} =0~ \forall j \in V_T,~ \nu\! \neq \!\mu \in V_B$. For instance, all stars with centers belonging to $V_B$ and with leaves of length $1$ or $2$ fall into this extended class.  Again, the retrieval of pattern $\mu$ and possibly of further patterns $\nu$ is achieved. However, the mutual signs of magnetizations are no longer arbitrary, as the terms $m_{\mu} \xi_i^{\mu} = m_{\nu} \xi_i^{\nu}$ are subject to constraints.

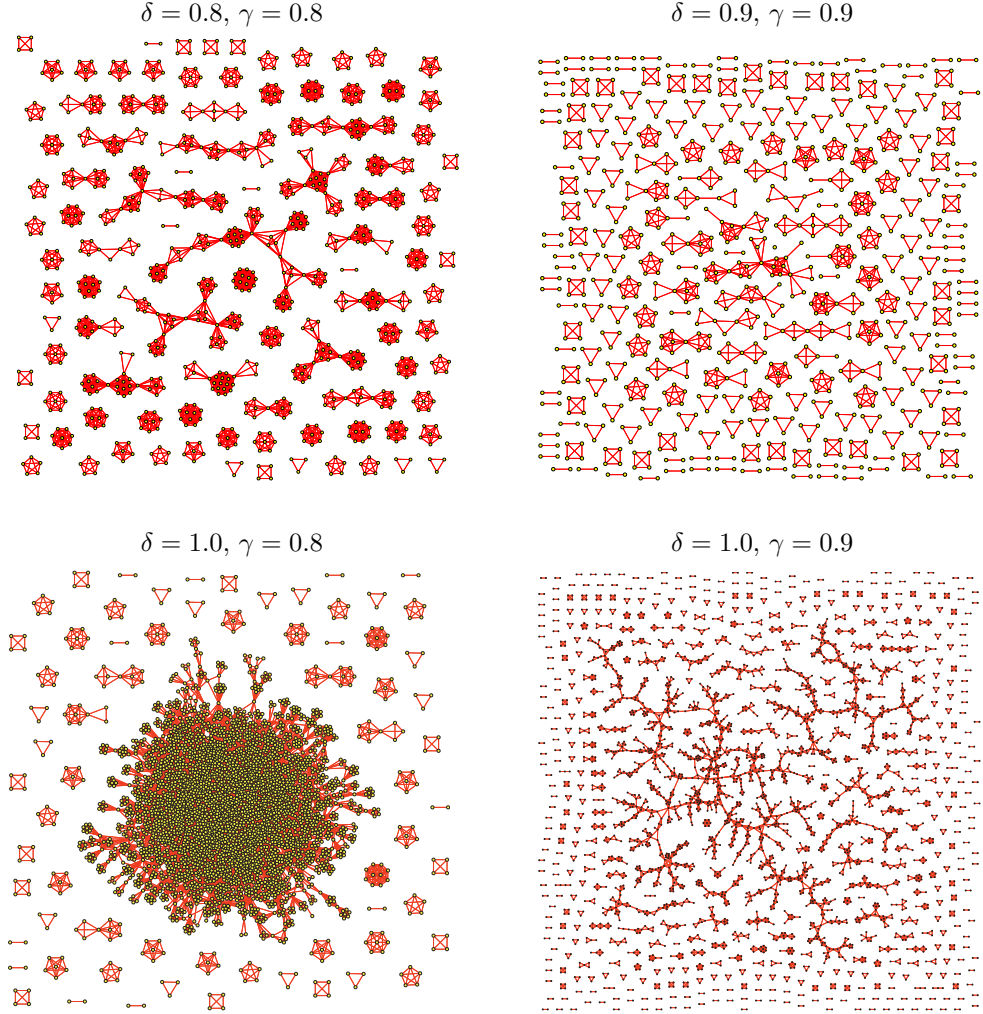
\begin{figure}[t]
\unitlength=0.235mm
\hspace*{22mm}
\begin{picture}(600,560)
\put(0,300){ \includegraphics[width=250\unitlength]{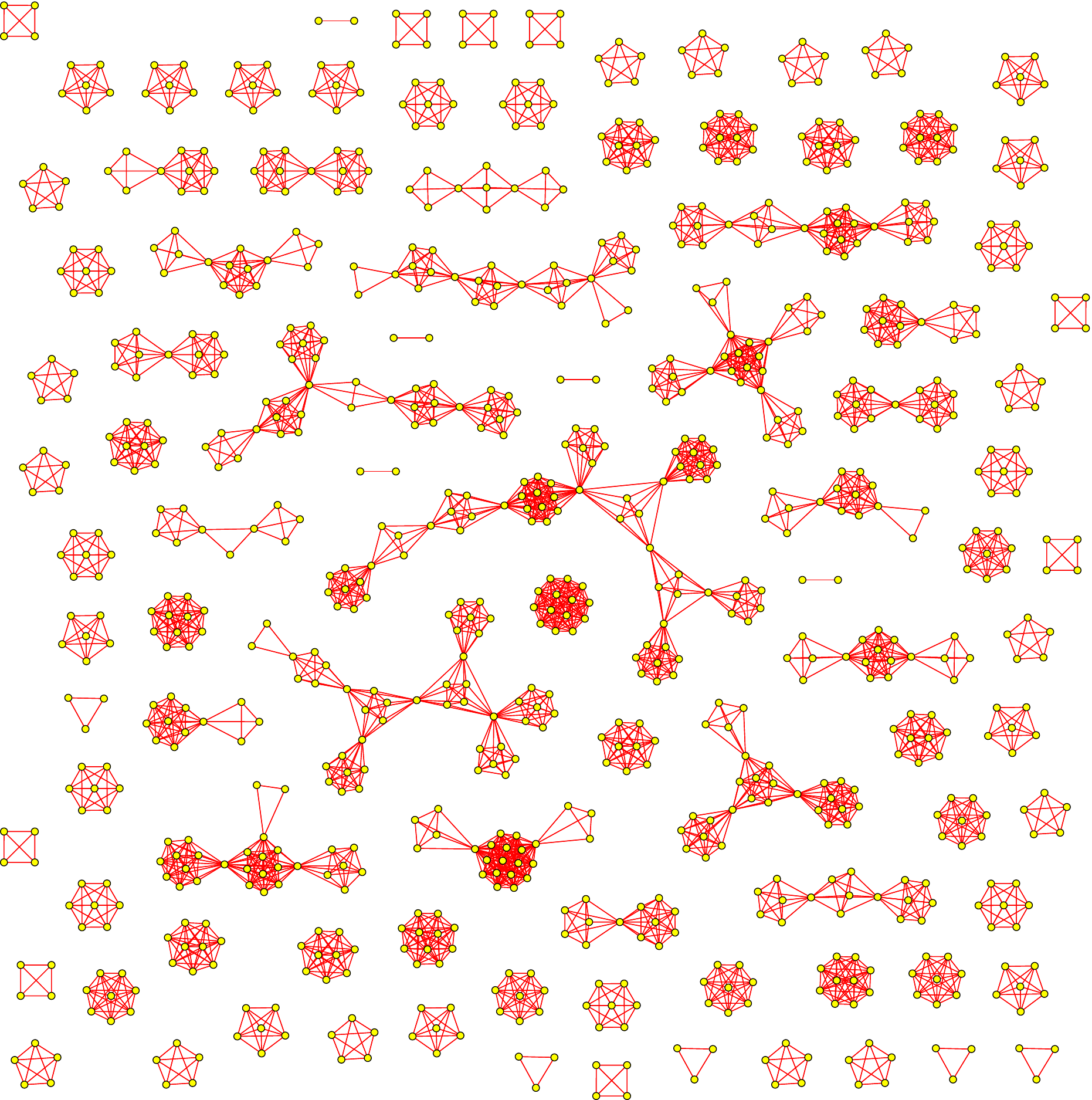}}
\put(70,560){ $\delta=0.8$, $\gamma=0.8$}
\put(300,300){\includegraphics[width=250\unitlength]{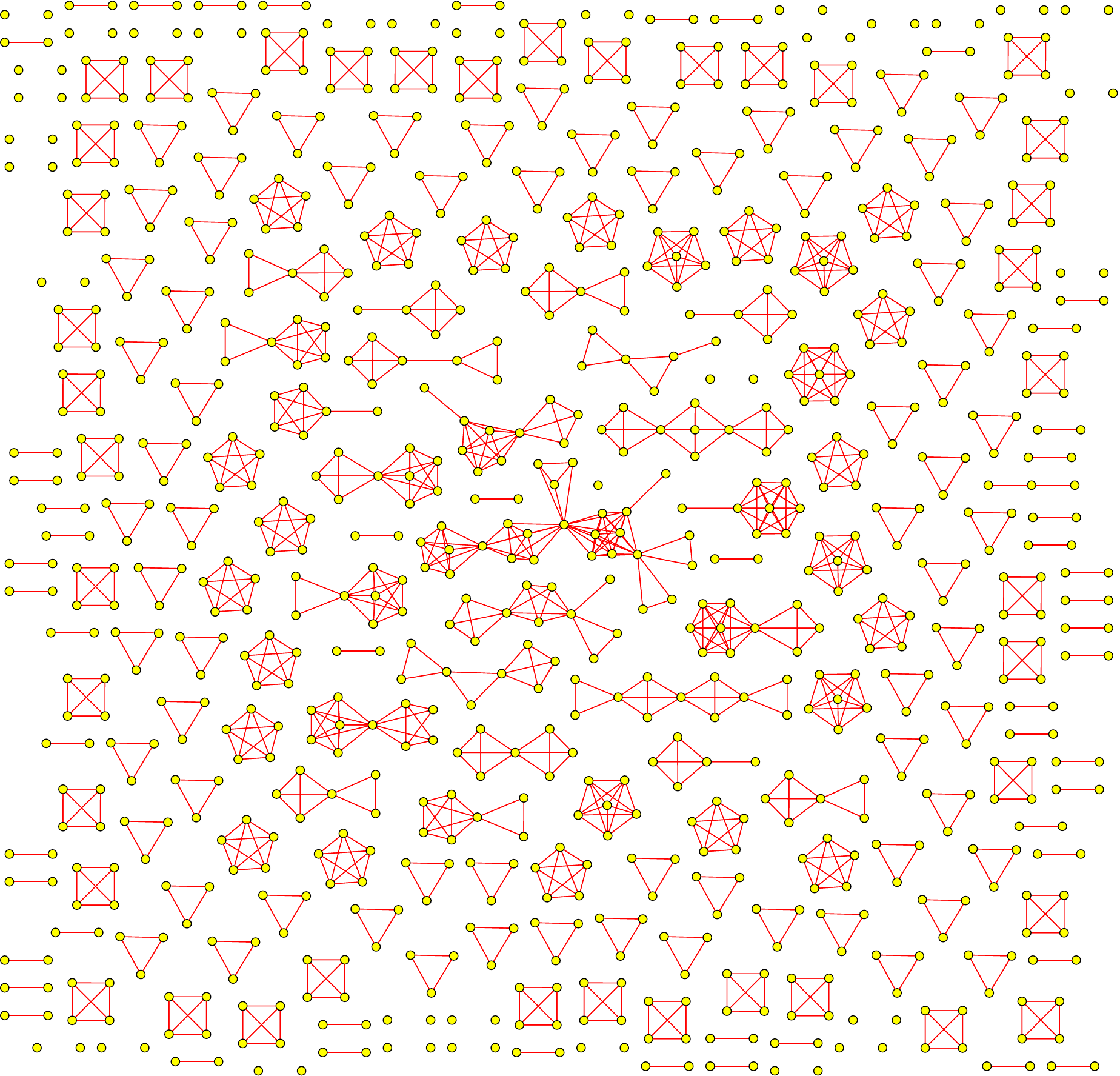}}
\put(370,560){ $\delta=0.9$, $\gamma=0.9$}
\put(0,0){\includegraphics[width=250\unitlength]{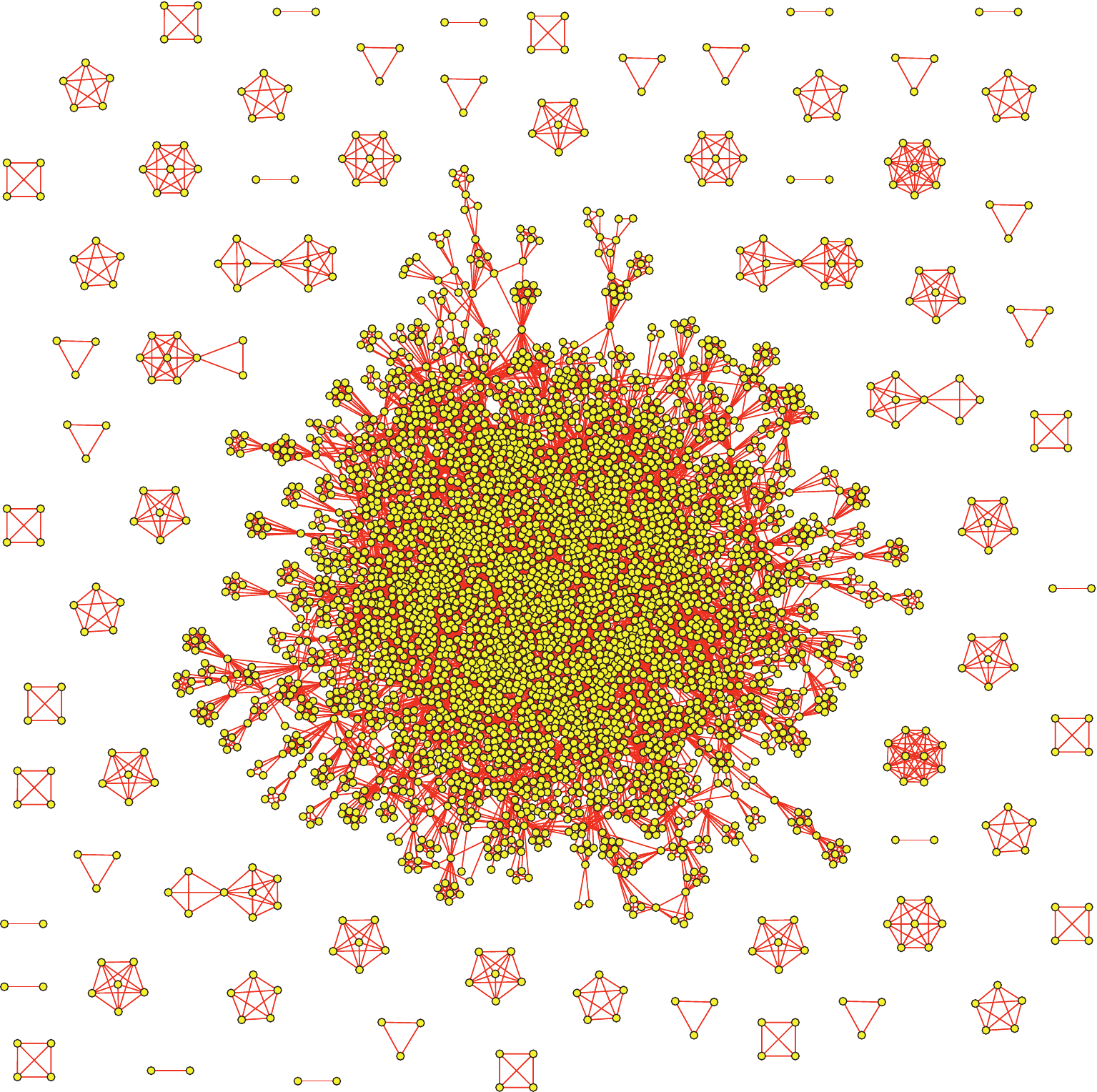}}
\put(70,260){ $\delta=1.0$, $\gamma=0.8$}
\put(300,0){\includegraphics[width=250\unitlength]{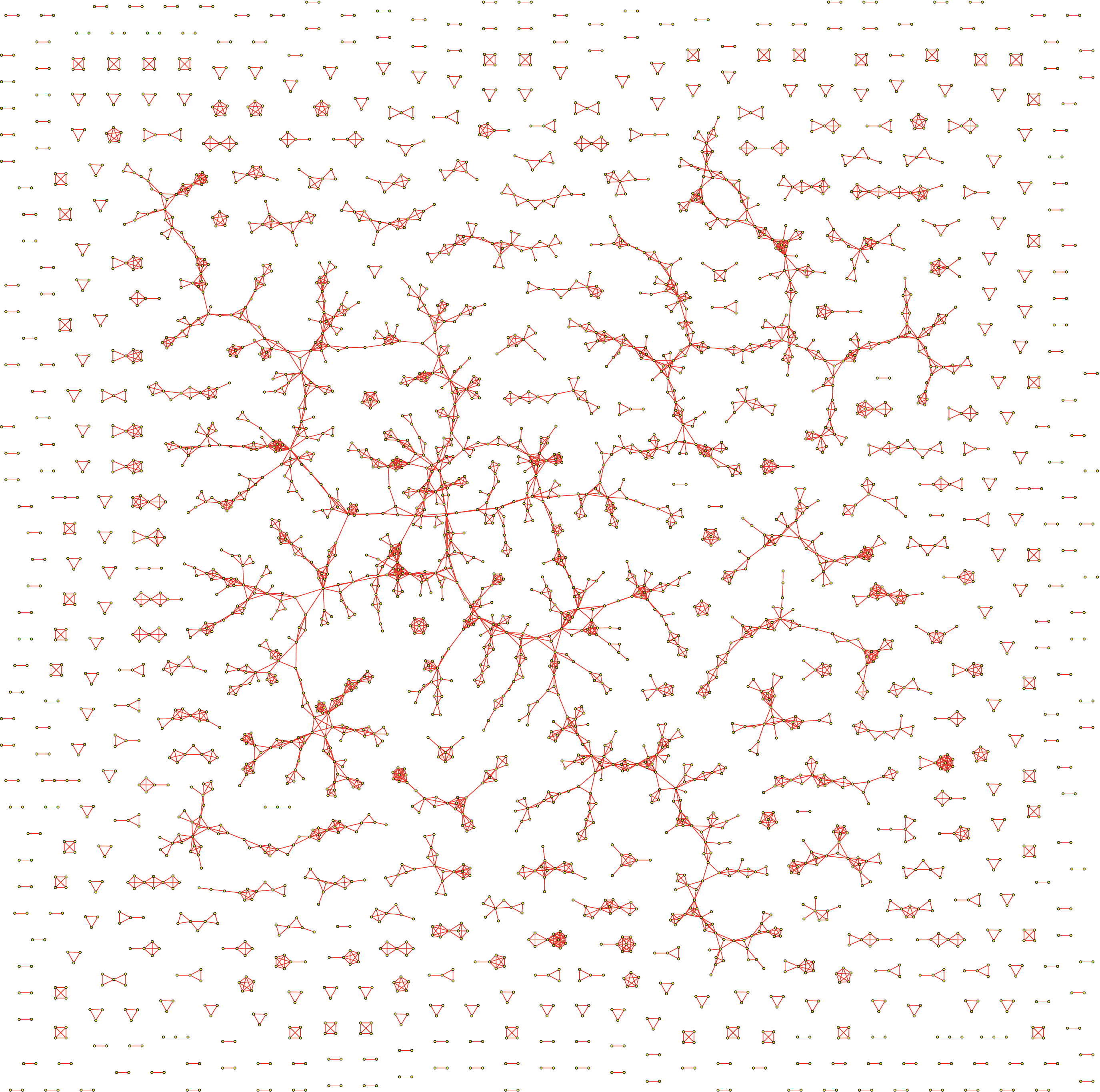}}
\put(370,260){ $\delta=1.0$, $\gamma=0.9$}
\end{picture}\vsp

\caption{\label{fig:grafi1} Plots of numerically generated graphs ${\mathcal G}$ for $N_T=10^4$ and $\alpha=0.1$, and different combinations of $(\delta,\gamma)$. All isolated nodes have been omitted from the plots. The chosen parameter combinations all give graphs that are just below the percolation threshold $\delta=2\gamma-1$. However, the two graphs at the top satisfy  the further condition $\delta\leq \gamma$ that marks the ability of simultaneous pattern retrieval (via weakly connected small cliques). In the  bottom graphs this latter condition is violated, so these would behave more like conventional Hopfield networks (here simultaneous retrieval of multiple patterns is not possible). }
\label{fig:grafi}
\end{figure}

We can further generalize the topology of components in $\mathcal{G}$ compatible with parallel retrieval, by considering cliques (i.e.  subsets of nodes such that every two nodes in the subset are connected by an edge), which are joined together by one link: each clique consists of  nodes $\in V_T$ that share the same non-null entry, so that the unique link between two cliques is due to a node displaying at least two non-null entries. This kind of structure exhibits a high degree of modularity; each clique is a module and corresponds to a different pattern. As for retrieval, this arrangement works fine as there is no interference between the signal on each node in $V_T$. For this arrangement to occur a sufficient condition is that $\mathcal{B}$ is devoid of squares, so that two nodes $\in V_T$ do not share more than one neighbor. This implies that, among the $n$ nodes connected to $j \in V_B$, the probability that any number $k>2$ of these display another common neighbor is vanishing:
\begin{equation}
\lim_{N_T\to\infty}\sum_{k>2}^{n} \Big(\!\!\begin{array}{c}n\\k \end{array}\!\!\Big)
p^k (1\!-\!p)^{n-k} = \lim_{N_T\to\infty} \Big\{1 -\left (1\!-\!\frac{c}{N^{\gamma}} \right)^n -\frac{cn}{N^{\gamma}} \left(1\!-\!\frac{c}{N^{\gamma}} \right)^{n-1}\Big\} =0.
\end{equation}
Since $n \sim N_T^{\d - \g}$, we obtain the condition $\g \geq \d$.
Examples of numerically generated graphs $\mathcal{G}$, for different choices of parameters, are shown in Fig.~\ref{fig:grafi}.

\subsection{Clustering properties}

\begin{figure}[t]
\unitlength=0.25mm
\hspace*{10mm}
\begin{picture}(600,345)
\put(0,0){ \includegraphics[width=600\unitlength]{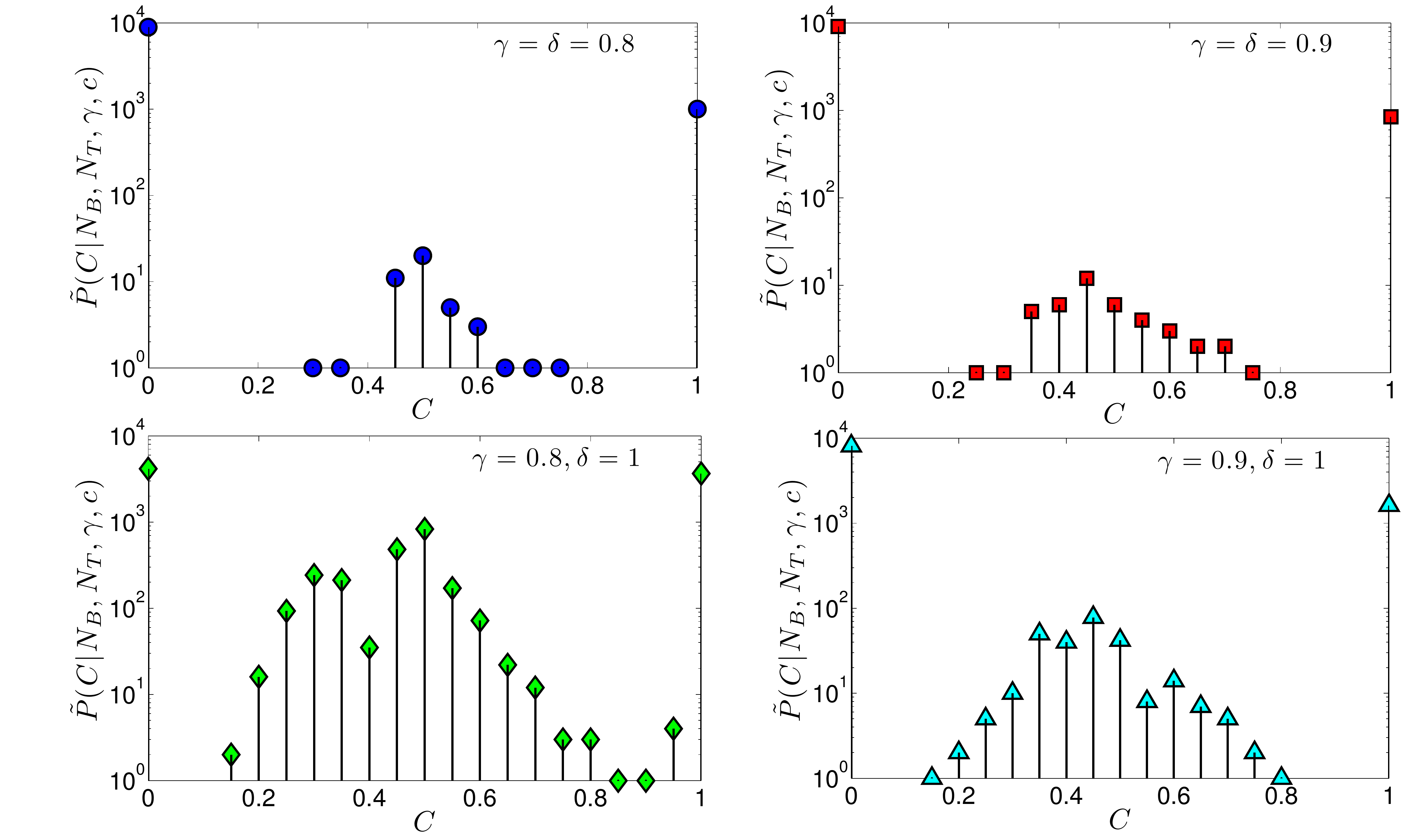}}
\end{picture}
\caption{\label{fig:clumod1} Histograms for $\tilde{P}(C | N_B, N_T, \gamma, c)$, normalized such that $\tilde{P}(C | N_B, N_T, \gamma, c) =1$ for those values of $C$ that have the lowest occurrence frequency.  This normalization allows to highlight the intermediate region, where the frequencies are much lower than those pertaining to $C=0$ and $C=1$.
The results shown here  refer to graphs with $N_T=10^4$ nodes, $\alpha=0.1$ and $c=1$, while $\delta$ and $\gamma$ are varied.}
\end{figure}

We saw that in the operationally most important parameter regime the graph $\mathcal{G}$ is built of small cliques which are poorly (if at all) connected to each other. This means that non-isolated nodes are highly clustered and the graph will have a high degree of modularity, i.e. dense connections between nodes within the same `module', but sparse connections between nodes in different `modules'.
The clustering coefficient $C_i$ of a node $i$ measures how close its $z_i$ neighbors  are to being a clique. It is defined as
\be
C_i = 2E_i/z_i (z_i-1),
\ee
where $E_i$ is the number of links directly connecting nodes pairs in $V_i$, while $\frac{1}{2}z_i(z_i-1)$
 is the total number of non-ordered node pairs in $V_i$. Hence $C_i\in[0,1]$.
The average clustering coefficient $\bar{C}=N_{B,T}^{-1}\sum_{i\in V_{B,T}}C_i$  measures the extent to which nodes in a graph tend to cluster together.
It is easy to see that for a bipartite graph, by construction, $C_i=0$ for any node, while for homogeneous graphs the local clustering coefficients are narrowly distributed around $\bar{C}$. For instance, for the Erd\"{o}s-R\'{e}ny random graph, where links are identically and independently drawn with probability $q$, the local coefficients are peaked around $q$. As for our graphs $\mathcal{G}$, due to their intrinsic inhomogeneity, the global measure $\bar{C}$ would give only limited information.  In contrast, the distribution $P(C| N_B, N_T, \gamma, c)$ of local clustering coefficients informs us about the existence and extent of cliques or `bulk' nodes, which would be markers of low and high recall interference, respectively.
Indeed, as shown in Figure~\ref{fig:clumod1}, in the highly-diluted regime most of the nodes in $\mathcal{G}$ are either highly clustered, i.e. exhibiting $C_i=1$, or isolated, with $C_i=0$, whereas the coefficients of the remaining nodes are  distributed around intermediate values with average decreasing with $\gamma$, as expected. In particular, when both $\delta$ and $\gamma$ are relatively large, $P(C | N_B, N_T, \gamma, c)$ approaches a bimodal distribution with peaks at $C=0$ and $C=1$, whereas when $\delta$ is sufficiently larger than $\gamma$, there exists a fraction of nodes with intermediate clustering which make up a bulk.
Therefore, although the density of links is rather small, the average clustering coefficient is very high,  and this is due to the fragmentation of the graph into many small cliques.

\begin{figure}[t]
 \begin{center}\vspace*{-2mm}\hspace*{5mm}
\includegraphics[width=360\unitlength]{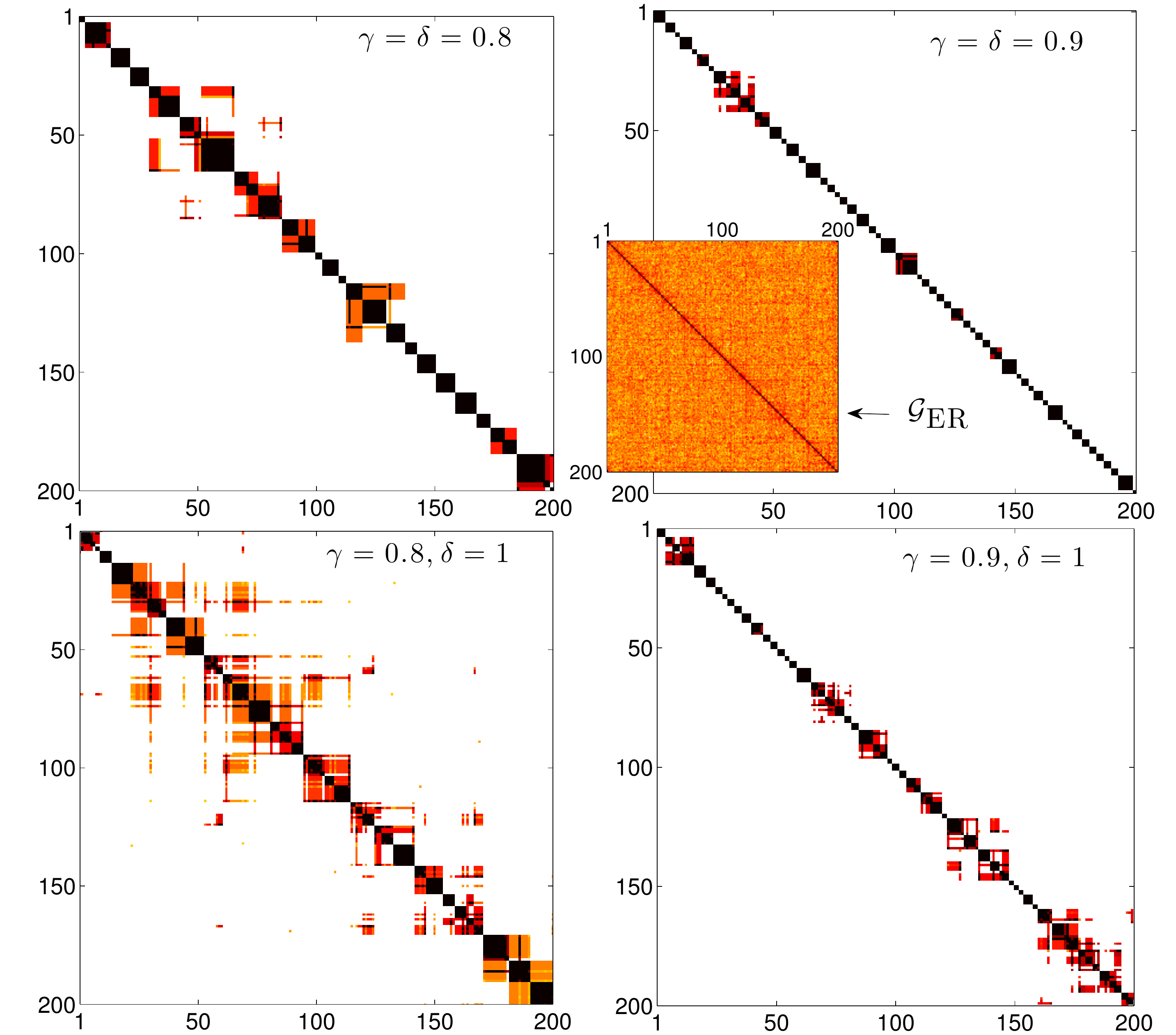}
\vsp

\caption{\label{fig:clumod2} Overlap matrix \textbf{T} with entries $T_{ij}=\sum_{k\neq i,j}c_{ik}c_{jk}/z_i$. The nodes are ordered such that nodes with large overlaps are adjacent, and the most significant part of \textbf{T}  is around the  diagonal. Note: $T_{ii}=1$ for all $i$, by construction. Darker colours correspond to larger entries, and any extended coloured zone denotes a module, i.e. a set of nodes that are highly clustered and possibly not connected with the remaining nodes. These plots refer to numerically generated graphs with $N_T=10^4$ nodes, $\alpha=0.1$ and $c=1$, while $\delta$ and $\gamma$ are varied. To avoid cluttering of the figures, only a fraction $200 \times 200$ of each pattern is shown. A similar fraction of the overlap matrix for an extremely diluted Erd\"{o}s-R\`enyi graph made of $10^4$ nodes is also depicted for comparison (inset). In this case there is no evidence of modularity; \textbf{T} displays a homogeneous pattern.}
\end{center}
\end{figure}

To measure the extent of modular structures we constructed the topological overlap matrix $\mathbf{T}$, whose entry $T_{ij}=\sum_{k\neq i,j}c_{ik}c_{jk}/z_i\in[0,1]$ returns the normalized number of neighbors that $i$ and $j$ share. The related patterns for several choices of parameters are shown in the plots of Fig.~\ref{fig:clumod2}, and compared  to those of Erd\"{o}s-R\`enyi graphs $\mathcal{G}_{\textrm{ER}}$. For Erd\"{o}s-R\`enyi graphs $\mathbf{T}$ displays a homogeneous pattern, that is very different from the highly clustered cases emerging from $\mathcal{G}$. In particular, for the highly diluted cases considered here, we find that smaller values of $\gamma$ induce a smaller number of modules, that are individually increasing in size.

\section{Medium storage regime in extremely diluted connectivity: retrieval region}

We now turn to the statistical mechanics analysis, and consider the immune network model composed of $N_T$ T-clones ($\s_i$, $i=1,\dots , N_T$) and $N_B$ B-clones ($b_{\mu}$, $\mu=1,\dots , N_B$), such that the number ratio scales as
\be
\lim_{N_T\to\infty}N_B/N_T^{\d}=\a,~~~~~~\d\in(0,1),~~ \alpha>0
\ee
The effective interactions in  the reduced network with helper cells only are described by the Hamiltonian
\be
\Hi(\bsigma|\xi)=-\frac{1}{2N_T^{\tau}}\sum_{i,j=1}^{N_T}\sum_{\mu=1}^{N_B}\xi^{\mu}_i\xi^{\mu}_j\s_i\s_j,
\label{eq:effHamiltonian}
\ee
where  the cytokine components $\xi_i^{\mu}\in \{0,\pm 1\}$ are
quenched  random variables, independently and identically distributed according to
\be
P(\xi^{\mu}_i=1)=P(\xi^{\mu}_i=-1)=c/2N_T^{\g},~~~~~~ P(\xi^{\mu}_i=0)=1-c/N_T^{\g}
\label{eq:pattern_stats}
\ee
with $\g\in[0,1)$. The parameter $\tau$ must be chosen such that $\Hi(\bsigma|\xi)$ scales linearly with $N_T$, and must therefore depend on $\g$ and $\d$.
Heuristically, since the number of non-zero entries $\Nu_{nz}$ in a generic pattern $(\xi_1^{\mu},\ldots,\xi_{N_T}^\mu)$ is $\order (N_T^{1-\g})$, we expect  that the network can retrieve a number of patterns of order  $\order(N_T/\Nu_{nz})=\order(N_T^{\g})$.
We therefore expect to see changes in $\tau$  only when crossing the region in the $(\gamma, \delta)$ plane where pattern sparseness prevails over storage load (i.e. $\d<\g$, where the system can recall all patterns), to the opposite situation, where the load is too high and frustration by multiple inputs on the same entry drives the network to saturation (i.e. $\d>\g$).
To validate this scenario, which is consistent with our previous topological investigation, we carry out a statistical mechanical analysis, based on computing the free energy
\be
f(\beta)=-\lim_{N_T\to\infty}\frac{1}{\b N_T} \langle \log Z_{N_T}(\b,\xi)\rangle_{\xi}.
\ee

\subsection{Free energy computation and physical meaning of the parameters}

If the number of patterns is sufficiently small compared to $N_T$, i.e. $\d<1$,
we do not need the replica method; we can simply apply the steepest descent technique using the $N_B \ll N_T/\log N_T$ Mattis magnetizations as order parameters:
\be
f(\beta)=-\frac 1 {\b} \log 2-\lim_{N_T \to \infty}\frac{1}{\b N_T}\log\int\! \rmd\bm\  e^{-\frac 1 2\bm^2+N_T\left\langle \log \cosh\left(\sqrt{\b/N_T^{\tau}}\bxi\cdot\bm\right)\right\rangle_{\xi}}.
\ee
with $\bm=(m_1,\ldots,m_{N_{B}})$, $\bxi=(\xi^1,\ldots,\xi^{N_{B}})$ and $\bxi\cdot\bm=\sum_\mu \xi^\mu m_\mu$. Rescaling of the order parameters via $m_{\mu}\to m_{\mu}\sqrt{\b}cN_T^{\tau/2+\theta}$ then gives
\bea
f(\beta)&=&-\frac 1 {\b} \log 2-\lim_{N_T \to \infty}\frac{1}{\b N_T}\log\int\! \rmd\bm\  e^{N_T \left(-\frac {\b c^2} 2 N_T^{\tau+2\theta-1} \bm^2 +\left\langle \log \cosh\left(\b c N_T^{\theta}\bxi\cdot\bm\right)\right\rangle_{\xi}\right)},
\eea
Hence, provided the limit exists, we may write via steepest descent integration:
\be
f(\beta)=-\frac 1 {\b} \log 2-\frac 1 {\b}\lim_{N_T \to \infty} {\rm extr}_{\bm}\left[\left\langle \log \cosh\left(\b c N_T^{\theta}\bxi\cdot\bm\right)\right\rangle_{\xi}-\frac {\b c^2} 2 N_T^{\tau+2\theta-1} \bm^2 \right].
\label{eq:free}
\ee
Differentiation with respect to the $m_{\mu}$ gives the self consistent equations for the extremum:
\be
\label{mselfc}
m_{\mu}=\frac{N_T^{1-\tau-\theta}}{c}\langle \xi^{\mu}\tanh(\b c N_T^{\theta}\bxi\cdot \bm)\rangle_{\xi}.
\ee
With the additional new parameter $\theta$,  we now have two parameters with which to control separately two types of normalization: the normalization of the Hamiltonian, via $\tau$, and the normalization of the order parameters, controlled by $\theta$. To carry out this task properly, we need to understand the physical meaning of the order parameters. This is done in the  usual way, by adding suitable external fields to the Hamiltonian:
\be
\Hi\to\Hi -\sum_{\mu=1}^{N_B}\l_{\mu}\sum_{i=1}^{N_T}\xi^{\mu}_i\s_i
\ee
Now, with $\langle g(\bsigma)\rangle_\sigma=Z^{-1}_{N_T}(\beta,\xi)\sum_{\bsigma}\rme^{-\beta \Hi(\bsigma|\xi)}g(\bsigma)$ and the corresponding new free energy $f(\beta,\blambda)$,
\be
\lim_{N_T\to\infty}\frac{1}{N_T} \langle \sum_{i=1}^{N_T}\xi^{\mu}_i\s_i\rangle_{\s}=-\frac{\partial f(\beta,\blambda)}{\partial \l_{\mu}}\Big|_{\blambda=0},
\label{eq:identify}
\ee
 with the short-hand $\blambda=(\lambda_1\ldots,\lambda_{N_B})$.
The new free energy is then found to be
\be
f(\beta,\blambda)=-\frac 1 {\b} \log 2-\frac 1 {\b}\!\lim_{N_T \to \infty} {\rm extr}_{\bm}\!\left[\left\langle \log \cosh\left(\b\bxi\cdot[ c N_T^{\theta}\bm\!+\!\blambda]\right)\right\rangle_{\xi}\!-\frac {\b c^2} 2 N_T^{\tau+2\theta-1} \bm^2 \right]~
\ee
 Upon differentiation with respect to
 $\l_{\mu}$ we find (\ref{eq:identify}) taking the form
\be
\lim_{N_T\to\infty}\frac{1}{N_T}\langle \sum_{i=1}^{N_T}\xi^{\mu}_i\s_i\rangle_{\s}
= \lim_{N_T\to \infty}\langle \xi^{\mu}\tanh(\b c N_T^{\theta}\bxi\cdot\bm)\rangle_{\xi}.
\ee
We can then use expression (\ref{mselfc}) for $m_{\mu}$ to obtain the physcal meaning of our order parameters:
\bea
m^{\mu}&=&\lim_{N_T\to\infty}\frac{N_T^{1-(\tau+\theta)}}{c}\langle \xi^{\mu}\tanh(\b c N_T^{\theta}\bxi\cdot\bm)\rangle_{\xi}\nn\\
&=&\lim_{N_T\to\infty}\langle\frac{1}{c N_T^{\tau+\theta}} \sum_{i=1}^{N_T}\xi^{\mu}_i\s_i\rangle_{\s}.
\eea
Let us summarize the status of the various remaining control parameters in the theory, in the interest of transparency.
Our model has three given external parameters:
\begin{itemize}
\item $\g\in[0,1)$: this quantifies the dilution of stored patterns,  via $P(\xi^{\mu}_i\neq 0)=c N_T^{-\g}$,
\item $\d\in(0,1)$ and $\a>0$: these determine the number of stored patterns, via $\lim_{N_T\to\infty}N_B/N_T^{\d}=\a$.
\end{itemize}
It also has two `internal' parameters, which must be set in such a way for the statistical mechanical calculation to be self-consistent, i.e. such that various quantities scale in the physically correct way for $N_T\to\infty$:
\begin{itemize}
\item $\tau\geq 0$: this must ensure  that the energy $\Hi=-\langle\frac{1}{2}N_T^{-\tau}\sum_{\mu=1}^{N_B}(\sum_{i=1}^{N_T} \xi^{\mu}_i\s_i)^2\rangle_{\sigma} $ scales as $\order(N_T)$,
\item $\theta\geq 0$: this must ensure that the order parameter $m_\mu=\left\langle (1/cN_T^{\tau+\theta})\sum_i\xi^{\mu}_i\s_i\right\rangle_\sigma$ are of order $\order(1)$.
\end{itemize}

\subsection{Setting of internal scaling parameters}

 To find the appropriate values for the internal scaling parameters $\theta$ and $\tau$ we return to the order parameter equation
(\ref{mselfc}) and carry out the average over $\xi^\mu$. This gives
\bea\label{selfcm}
m_{\mu}&=&\frac{N_T^{1-\tau-\theta}}{c}\langle \tanh\Big(
\b c N_T^{\theta}\Big((\xi^\mu)^2 m_\mu
+\xi^{\mu}
\sum_{\nu\neq \mu}^{N_B}\xi^\nu m_\nu\Big)
\Big)\rangle_{\xi}.
\\
&=& N_T^{1-\tau-\theta-\gamma}\langle \tanh\big(
\b c N_T^{\theta}\big(m_\mu
+
\sum_{\nu\neq \mu}^{N_B}\xi^\nu m_\nu\big)
\big)\rangle_{\xi}.
\eea
Having non-vanishing $m_\mu$ in the limit $N_T\to\infty$ clearly demands $\theta+\tau\leq 1-\gamma$. If $\theta>0$
the $m_\mu$ will become independent of $\beta$, which means that any phase transitions occur ar zero or infinite noise levels, i.e. we would not have defined the scaling of our Hamiltonian correctly. Similarly, if $\theta+\tau< 1-\gamma$ the effective local fields acting upon the $\sigma_i$ (viz. the arguments of the hyperbolic tangent) and therefore also the expectation values $\langle \sigma_i\rangle_\sigma$, would be vanishingly weak. We therefore conclude that a natural ansatz for the free exponents is:
\begin{eqnarray}
(\tau,\theta)=(1-\gamma,0)
\end{eqnarray}
This
simplifies the order parameter equation to
\bea
m_{\mu}&=&\langle \tanh\big(
\b c \big(m_\mu
+
\sum_{\nu\neq \mu}^{N_B}\xi^\nu m_\nu\big)
\big)\rangle_{\xi}.
\eea
Let us analyze this equation further.  Since $P(\xi^{\mu}_i\neq 0)\sim N_T^{-\g}$ with $\g>0$, we can for $N_T\to\infty$ replace in $(\ref{selfcm})$ the sum over $\nu\neq\mu$ with the sum over all $\mu$; the difference is negligible in the thermodynamic limit. In this way it becomes clear that   for each solution of $(\ref{selfcm})$ we have $m_{\mu}\in \{-m,0,m\}$. Using the invariance of the free energy under $m_\mu\to -m_\mu$,  we can from now on focus on solutions with non-negative magnetizations. If we denote with $K\leq N_B$ the number of $\mu$ with $m_\mu\neq 0$, then
the value of $m>0$ is to be solved from
\bea
m&=&\langle \tanh\big(
\b c m\big(1
+
\sum_{\nu=1}^K\xi^\nu\big)
\big)\rangle_{\xi}.
\label{eq:findm}
\eea
It is not a priori obvious how the number $K$ of nonzero magnetizations (i.e. the number of simultaneously triggered clones) can or will scale with $N_T$. We therefore set
$K=\phi N_T^{\d^\prime}$, in which the condition $K\leq N_B$ then places the following conditions on $\phi$ and $\delta^\prime$:
 $\d^\prime\in [0,\d]$, and  $\phi\in[0,\infty)$ if $\d^\prime<\d$ or $\phi\in[0,\a]$ if $\d^\prime=\d$.
We expect that if $K$ is too large, equation (\ref{eq:findm}) will only have the trivial solution for $N_T\to \infty$, so there will be further conditions on $\phi$ and $\delta^\prime$ for the system to operate properly.
If $\d'>\g$, the noise due to other condensed patterns (i.e. the sum over $\nu$) becomes too high, and $m$ can only be zero:
\be
\E \left[ \big(\sum_{\mu=1}^K\xi^{\mu}\big)^2\right] =\sum_{\mu=1}^K\E[{\xi^{\mu}}^2]=\phi c \frac{N_T^{\d'}}{N_T^{\g}}\to\infty.
\ee
On the other hand, if $\d '<\g$ this noise becomes negligible,  and (\ref{eq:findm}) reduces to the Curie-Weiss equation, whose solution is just the Mattis magnetization \cite{ton0,amit,BarraCW}. It follows that the
 critical case is the one where when $\d^\prime=\g$. Here  we have for $N_T\to\infty$ the following equation for $m$:
\begin{eqnarray}
m&=&\sum_{k\in \Z} \pi(k|\phi)\tanh(\beta c m(1+k))
\label{eq:minpi}
\end{eqnarray}
with the following discrete noise distribution, which obeys $\pi(-k|\phi)=\pi(k|\phi)$:
\begin{eqnarray}
\pi(k|\phi)&=&\big\langle \delta_{k,\sum_{\mu=1}^\infty\xi^{\mu}}\big\rangle_\xi
\end{eqnarray}

\subsection{Computation of the noise distribution $\pi(k)$}

Given its symmetry, we only need to calculate $\pi(k|\phi)$ for $k\geq 0$:
\bea
\pi(k|\phi)
&=&\lim_{K\to\infty} \int_{-\pi}^{\pi}\frac{\rmd\psi}{2\pi}e^{-i\psi k}\left\langle e^{i\psi\xi}\right\rangle_{\xi}^{K}
=\lim_{K\to\infty}\int_{-\pi}^{\pi}\frac{\rmd\psi}{2\pi}e^{-i\psi k}\big(1+\frac {c\phi} {K}(\cos\psi-1)\big)^{K}\nn\\
&=&\int_{-\pi}^{\pi}\frac{\rmd\psi}{2\pi}e^{-i\psi k+\phi c (\cos\psi -1)}
\nn
\\
&=&e^{-\phi c}\int_{-\pi}^{\pi}\frac{\rmd\psi}{2\pi}e^{-i\psi k}\sum_{n\geq 0}\frac{(\phi c)^n}{2^n n!}(e^{i\psi}+e^{-i\psi})^n\nn\\
&=&e^{-\phi c}\int_{-\pi}^{\pi}\frac{\rmd\psi}{2\pi}e^{-i\psi k}\sum_{n\geq 0}\frac{(\phi c)^n}{2^n n!}\sum_{l\leq n}\frac{n!}{l!(n-l)!}e^{-i\psi(k-n+2l)}\nn\\
&=& e^{-\phi c}\sum_{n\geq 0}\sum_{l\leq n}\left(\frac{\phi c}{2}\right)^n\!\!\frac{1}{l!(n-l)!}\delta_{n,k+2l}\nn\\
&=& e^{-\phi c}\sum_{l\geq 0}\left(\frac{\phi c}{2}\right)^{k+2l}\!\!\frac{1}{l!(k+l)!}~=~e^{-\phi c}~\mathcal{I}_k(\phi c)
\label{eq:pik}
\eea
where $\mathcal{I}_k(x)$ is the $k$-th modified Bessel function of the first kind \cite{AbramSteg}. These modified Bessel  functions obey
\bea
2\frac{k}{x}\mathcal{I}_k(x)&=&\mathcal{I}_{k-1}(x)-\mathcal{I}_{k+1}(x),\nn\\
2\frac{\rmd}{\rmd x}\mathcal{I}_k(x)&=&\mathcal{I}_{k-1}(x)+\mathcal{I}_{k+1}(x).
\eea
The first identity leads to a useful
 recursive equation for $\pi(k|\phi)$, and the second identity simplifies our calculation of derivatives of $\pi(k|\phi)$ with respect to $\phi$, respectively:
\begin{eqnarray}
\label{vinc}
&&\pi(k\!-\!1|\phi)-\pi(k\!+\!1|\phi)-2\pi(k|\phi)\frac{k}{\phi c}=0,
\\
&&\frac{\rmd}{\rmd\phi}\pi(k|\phi)= c\Big(\frac{1}{2}\pi(k\!-\!1|\phi)+\frac{1}{2}\pi(k\!+\!1|\phi)-\pi(k|\phi)\Big)
\label{eq:pider}
\end{eqnarray}

\subsection{Retrieval in the zero noise limit}

To emphasize the dependence of the recall overlap on $\phi$, viz. the relative storage load, we will from now on write $m\to m_\phi$.
With the abbreviation $\langle g(k)\rangle_k=\sum_k \pi(k|\phi)g(k)$, and using (\ref{vinc}) and the symmetry of $\pi(k|\phi)$, we can transfer our equation (\ref{eq:minpi}) into a more convenient form:
\begin{eqnarray}
m_\phi&=& \frac{1}{2}\langle\Big[\tanh(\beta c m_\phi(1\!+\!k))+\tanh(\beta c m_\phi(1\!-\!k))\Big]\rangle_k
\nn
\\
&=& \frac{1}{2}\sum_{k\in \Z} \Big[\pi(k\!-\!1|\phi)-\pi(k\!+\!1|\phi)\Big]\tanh(\beta c m_\phi k) =
\frac{1}{\phi c}\langle k\tanh(\beta c m_\phi k)\rangle_k
\label{eq:m_conv}
\end{eqnarray}
In the zero noise limit $\beta\to\infty$, where $\tanh(\beta y)\to {\rm sgn}(y)$, this reduces to
$m_\phi= \frac{1}{\phi c}\langle |k|\rangle_k$,
or, equivalently,
\bea
m_\phi&=&\lim_{\beta\to \infty}\langle \tanh(\beta c m(1\!+\!k))\rangle_k=\langle {\rm sign}(1\!+\!k)\rangle_k
\nn\\
&=&\sum_{k>-1} \!\pi(k)-\!\sum_{k<-1}\!\pi(k) ~=~\pi(0|\phi)+\pi(1|\phi),
\eea
Hence we always have a nonzero rescaled magnetization, for any relative storage load $\phi$. To determine for which value of $\phi$ this state is most stable, we have to insert this solution into the zero temperature formula for the free energy and find the minimum with respect to $\phi$.
Here, with $m_\mu=m_\phi$ for all $\mu\leq K=\phi N_T^\gamma$ and $m_\mu=0$ for $\mu>K$, the free energy (\ref{eq:free}) takes asymptotically the form
\be
f(\beta)=
\frac{1}{2} c^2\phi m_\phi^2
-\frac 1 {\b}
\langle \log \cosh\left(\b c m_\phi k\right)\rangle_{k}   -\frac 1 {\b} \log 2
\ee
So for $\beta\to \infty$, and using our above identity $\langle |k|\rangle_k=\phi c m$, we find that the  energy density is
\begin{eqnarray}
u(\phi)&=&\lim_{\beta\to\infty}
f(\beta)=
\frac{1}{2} c^2\phi m_\phi^2
-c m
\langle |k|\rangle_{k}   =
-\frac{1}{2} c^2\phi m_\phi^2
\\
&=& -\frac{1}{2} c^2\phi \big(\pi(0|\phi)\!+\!\pi(1|\phi)\big)^2
\end{eqnarray}
To see how this depends on $\phi$ we may use (\ref{eq:pider}), and find
\begin{eqnarray}
\frac{1}{c^2}
\frac{\rmd}{\rmd\phi}u(\phi)&=&
-\frac{1}{2} m_\phi^2
-\phi m_\phi \frac{\rmd}{\rmd \phi}\big(\pi(0|\phi)\!+\!\pi(1|\phi)\big)
\nonumber
\\
&=& -\frac{1}{2} m_\phi^2
-\phi cm_\phi \Big(
-\frac{1}{2}\pi(0|\phi)
+\frac{1}{2}\pi(2|\phi)\Big)
= -\frac{1}{2} m_\phi^2
+ m_\phi \pi(1|\phi)
\nonumber
\\
&=&  -\frac{1}{2} m_\phi\Big(m_\phi-2
\pi(1|\phi)\Big)=-\frac{1}{2} m_\phi\Big(\pi(0|\phi)-
\pi(1|\phi)\Big)<0
\end{eqnarray}
The energy density $u(\phi)$ is apparently a decreasing function of $\phi$, which reaches its minimum when the number of condensed patterns is maximal, at $\phi=\alpha$.
 However, the amplitude of each recalled pattern will also decrease for larger  values of $\phi$:
\begin{eqnarray}
\frac{\rmd}{\rmd\phi}m_\phi&=& \frac{\rmd}{\rmd\phi}\pi(0|\phi)+ \frac{\rmd}{\rmd\phi}\pi(1|\phi)
= -\pi(1|\phi)/\phi <0
\end{eqnarray}
Hence $m_\phi$ starts at $m_0=1$, due to $\pi(k|0)=\delta_{k,0}$, and then decays monotonically with $\phi$.
Moreover, it follows from $\langle|k|\rangle_k^2\leq \langle k^2\rangle_k=\langle \sum_{\mu\leq K}(\xi^\mu)^2\rangle_\xi=\phi c$ that
\begin{eqnarray}
m_\phi=\langle|k|\rangle_k/\phi c~\leq~ 1/\sqrt{\phi c},~~~~~~~~
u(\phi)= -\frac{1}{2}c^2\phi m_\phi^2~\geq~ -\frac{1}{2}c
\end{eqnarray}
If we increase the number of condensed patterns, the corresponding magnetizations decrease in  such a way  that the energy density remains finite.

\subsection{Retrieval at nonzero noise levels}

\begin{figure}[t]
\vspace*{-2mm}
\centering
\includegraphics[width=120mm]{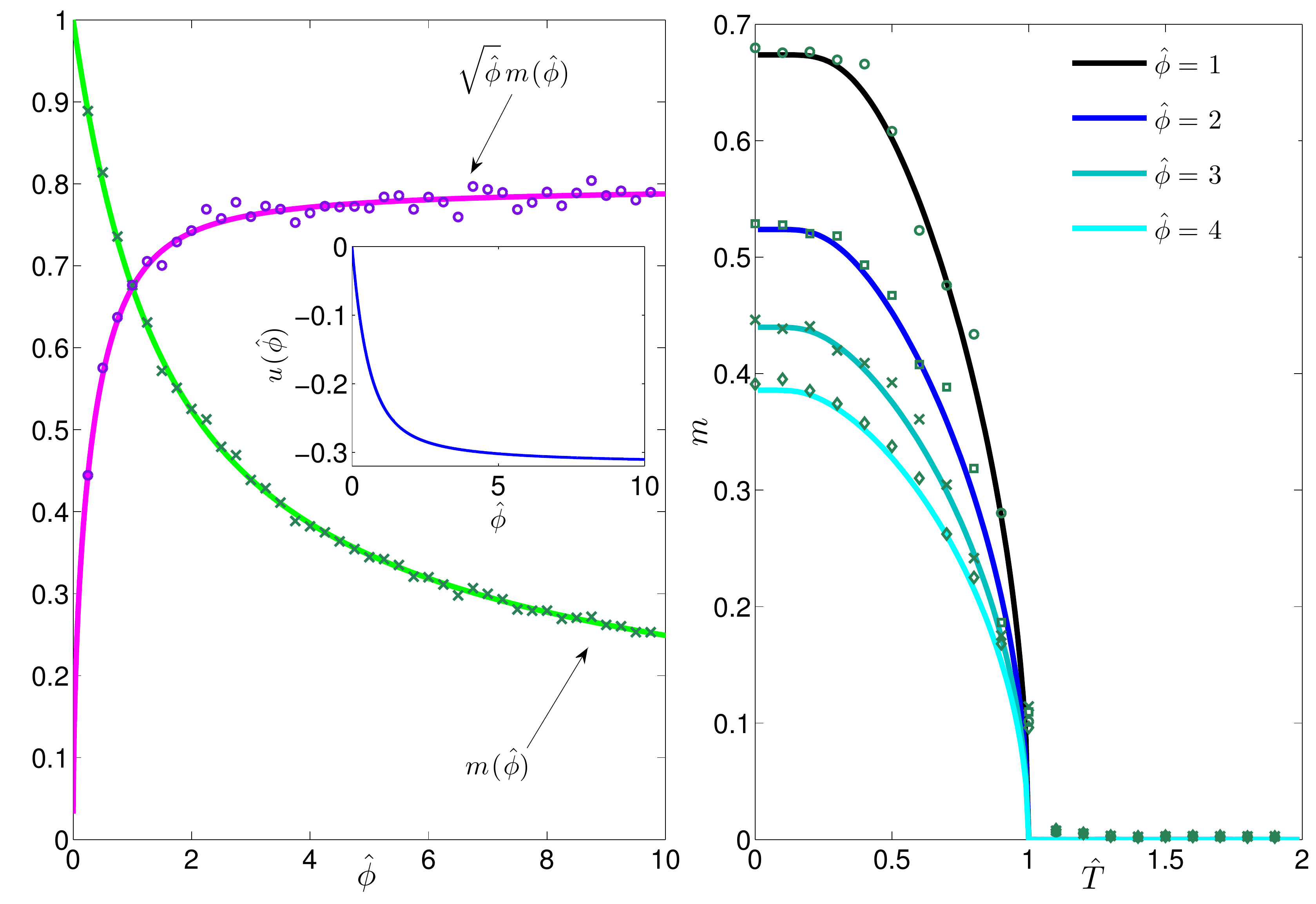}
\caption{Left: energy density $u$ versus the relative fraction of retrieved patterns, in terms of $\hat{\phi}=c\phi$ and $\hat{T}=T/c=1/\beta c$. The minimum energy density is reached when $\hat{\phi}$ is maximal, i.e. when {\em all} stored patterns are simultaneously  retrieved, but with decreasing amplitude for each. Right: critical noise levels for different values of $\hat{\phi}$, confirming that  $\hat{T}_c^{-1}=\hat{\beta}_c=1$, independently of $\hat{\phi}$. In both the panels, solid lines represent our theoretical predictions,  while symbols represent data from numerical simulations on systems with $N_T=5\times10^4$, $\g=\d=0.45$, $c=2$ and with with standard sequential Glauber dynamics.}
\label{fig:TC}
\end{figure}

To find the critical noise level (if any) where pattern recall sets in, we return to equation (\ref{mselfc}), which for $(\tau,\theta)=(1\!-\!\gamma,0)$ and written in vector notation becomes
\be
\bm=\frac{N_T^{\gamma}}{c}\langle \bxi\tanh(\b c \bxi\cdot \bm)\rangle_{\xi}.
\ee
We take the inner product on both sides with $\bm$ and obtain a simple inequality:
\begin{eqnarray}
\bm^2&=&\frac{N_T^{\gamma}}{c}\langle (\bxi\cdot\bm)\tanh(\b c \bxi\cdot \bm)\rangle_{\xi}
\nonumber
\\
&=& \beta N_T^{\gamma}\langle (\bxi\cdot\bm)^2\int_0^1\!\rmd x~[1-\tanh^2(\b c x\bxi\cdot \bm)]\rangle_{\xi}
\nonumber
\\
&\leq & \beta N_T^{\gamma}\langle (\bxi\cdot\bm)^2\rangle_{\xi}
=\beta c \bm^2
\end{eqnarray}
Since $\bm^2(1-\beta c)\leq 0$, we are sure that $\bm=0$ for $\beta c\leq 1$. At $\beta c=1$ nontrivial solutions of the previously studied symmetric type are found to bifurcate continuously from the trivial solution.
This can be seen by expanding the amplitude equation (\ref{eq:m_conv}) for small $m$:
\begin{eqnarray}
m_\phi&=&
\frac{1}{\phi c}\langle k\tanh(\beta c m_\phi k)\rangle_k
\nonumber
\\
&=& \beta c m_\phi
-\frac{1}{3}\beta^3 c^2 m^3_\phi
 \langle k^4\rangle_k/\phi+\order(m_\phi^4)
\end{eqnarray}
This shows that the symmetric solutions indeed bifurcate  via  a second-order transition, at the $\phi$-independent critical temperature $T_c=c$, with amplitude $m_\phi \propto (\b c -1)^{\frac 1 2}$ as $\beta c\to 1$.
All the above predictions are confirmed by the results of numerical simulations, and by  solving the order parameter equations and calculating the free energy numerically, see
Figure \ref{fig:TC}.
\vsp

We can now summarize the phase diagram in terms of the scaling exponents ($\g,\d$). The number of stored patterns is $N_B=\alpha N_T^\delta$, of which $K=\phi N_T^{\delta^\prime}$ can be recalled simultaneously, with $\delta^\prime={\rm min}(\gamma,\delta)$:
\begin{description}
\item[~~~~$\d<\g\!\!$]: ~~$\phi_{\rm max}=\a$,~~ all stored patterns recalled simultaneously, with Curie-Weiss overlap $m$
\item[~~~~$\d=\g\!\!$]: ~~$\phi_{\rm max}=\a$,~~ all stored patterns  recalled simultaneously, with reduced but finite $m$
\item[~~~~$\d>\g\!\!$]: ~~$\phi_{\rm max}=\infty$, ~at most $\phi N_T^{\g}$ patterns recalled simultaneously,  with $\phi\to\infty$ and $m_{\phi}\to 0$
\end{description}

\section{High storage regime in extremely diluted connectivity: absence of retrieval}

Let us finally consider the same network, composed of $N_T$ T-clones ($\s_i$, $i=1,\dots , N_T$) and $N_B$ B-clones ($b_{\mu}$, $\mu=1,\dots , N_B$), but now at high storage load:
\be
\lim_{N_T\to\infty}N_B/N_T=\a, ~~~~~\alpha>0
\ee
The effective interaction between T-cells is still described by the Hamiltonian (\ref{eq:effHamiltonian}),
and the cytokine variables  $\xi_i^{\mu}\in \{0,\pm 1\}$  are generated from (\ref{eq:pattern_stats}),
but now we focus  on the extremely diluted regime for the B-T network, i.e. $\g<1$.
Again we must choose $\tau$ such that the Hamiltonian will be of order $N_T$. Heuristically, since the number of non-zero entries $\Nu_{\rm nz}$ in a typical pattern $(\xi_1^{\mu},\ldots,\xi_{N_T}^\mu)$ scales as $\order(N_T^{1-\g})$, the number of patterns with non overlapping entries (i.e. those we expect to recall) will scale as  $\order(N_T/\Nu_{nz})=\order(N_T^{\g})$. The contribution from $K=\order(N_T^{\g})$ such condensed patterns  to the Hamiltonian would then scale as
\be\nn
\Hi_C\sim N_T^{-\tau}\sum_{\mu=1}^{K} (\sum_{i=1}^{N_T} \xi^{\mu}_i\s_i)^2 \sim N_T^{-\tau} K \Nu_{\rm nz}^2\sim N_T^{-\tau}N_T^{\g}N_T^{2(1-\g)}\sim N_T^{2-\g-\tau}
\ee
The non-condensed patterns, of which there are $N_{\rm nc}=N_B-N_c\sim N_B=\order(N_T)$, are expected to contribute
\be\nn
\Hi_{NC}\sim N_T^{-\tau}\sum_{\mu=1}^{N_{\rm nc}} (\sum_{i=1}^{N_T} \xi^{\mu}_i\s_i)^2 \sim N_T^{-\tau} N_{\rm nc} \sqrt{\Nu_{nz}}^2\sim N_T^{-\tau}N_T N_T^{1-\g}\sim N_T^{2-\g-\tau}.
\ee
Thus, we expect to have an extensive Hamiltonian for $\tau=1-\g$.

\subsection{Replica-symmetric theory}

In the scaling regime $N_B=\alpha N_T$ we can no longer use saddle-point arguments directly in the calculation of the free energy. Instead we calculate the free energy  for typical cytokine realizations, i.e. the average
\be
\overline{f}=-\lim_{N_T\to\infty}\frac{1}{\b N_T} \overline{\log Z_{N_T}(\b,\xi)},
\ee
Here $\overline{\cdots}$ indicates averaging over all $\{\xi_i^\mu\}$, according to the measure
 (\ref{eq:pattern_stats}). The average over cytokine variables  is done with the replica method, for $K=\order(N_T^\gamma)$; full details are given in Appendix B. We solve the model at the replica symmetric (RS) level, which implies the assumption that the system has at most a finite number of ergodic sectors for $N_T\to\infty$, giving
\begin{eqnarray}
\b \overline{f}_{\rm RS}&=& \lim_{N_T\to \infty} \hbox{extr}_{\bm,q,r} ~\b\hat{f}_{\rm RS}(\bm,q,r),
\\
\beta\hat{f}_{\rm RS}(\bm,q,r)&=& -\log 2
+ \frac{1}{2}\a r(\b c)^2 (1\!-\!q)
+\frac{\beta c^2}{2N_T^\gamma}\bm^2
- \frac{\alpha}{2}\Big(\frac{\beta c q}{1\!-\!\beta c(1\!-\!q)}
-\log[1\!-\!\beta c(1\!-\!q)]\Big)
\nonumber
\\&&
- \Big\langle \int\!{\rm D}z ~\log
\cosh[ \b c(\bm\cdot\bxi
+z\sqrt{\alpha r})]
\Big\rangle_{\xi}
\eea
in which $\bm=(m_1,\ldots,m_K)$ denotes the vector of $K=\phi N_T^\gamma$ condensed (i.e. potentially recalled) patterns, $\bxi=(\xi^1,\ldots,\xi^K)$, and ${\rm D}z=(2\pi)^{-1/2}\rme^{-z^2/2}\rmd z$.
As in the analysis of standard Hopfield networks, this involves the Edward-Anderson spin-glass order parameter $q$ \cite{ton0,amit} and  the Amit-Gutfreund-Sompolinsky uncondensed-noise order parameter $r$ \cite{ton0,amit}. We obtain self-consistent equations for the remaining RS order parameters $(m,q,r)$ simply by extremizing $\hat{f}_{\rm RS}(\bm,q,r)$, which leads to
\bea\label{rssce}
m^{\mu}&=&\frac{N_T^{\g}}{c}\Big\langle \xi^{\mu}\!\int\!{\rm D}z ~\tanh[\b c(\bm\cdot\bxi\!+\!z\sqrt{\a r})]\Big\rangle_{\xi},\nn\\
q&=&\Big\langle\int\!{\rm D}z ~\tanh^2[\beta c(\bm\cdot\bxi\!+\!z\sqrt{\a r})]\Big\rangle_{\xi},\nn\\
r&=&\frac{q}{[1\!-\!\beta c(1\!-\!q)]^2}.
\eea
As before we  deal with the equation for $m^\mu$ by using the identity $\xi^\mu\tanh(A)=\tanh(\xi^\mu A)$ (since $\xi^\mu\in\{-1,0,1\}$) and by separating the term $m^\mu\xi^\mu$ from the sum $\bm\cdot\bxi$:
\bea\label{sp}
m^{\mu}&=&
\frac{N_T^\gamma}{c}\Big\langle \int\! {\rm D}z ~\tanh[\beta c(m^{\mu}(\xi^{\mu})^2+\sum_{\nu\neq\mu\leq K}m^{\nu}\xi^{\nu}\xi^\mu+z\xi^\mu\sqrt{\a r})]\Big\rangle_{\xi}\nn\\
&=&
\Big\langle \int\! {\rm D}z ~\tanh[\beta c(m^{\mu}+\sum_{\nu\neq\mu\leq K}m^{\nu}\xi^{\nu}+z\sqrt{\a r})]\Big\rangle_{\xi}\nn\\
&=&
\Big\langle \int\! {\rm D}z ~\tanh\Big[\beta c\Big(m^{\mu}+\sum_{\nu=1}^Km^{\nu}\xi^{\nu}+z\sqrt{\a r}\Big)\Big]\Big\rangle_{\xi}+\order(N^{-\gamma})\nn
\eea
Again we see that for $N_T\to\infty$ we will only retain solutions with $m^{\mu}\in \{-m,0,m\}$ for all $\mu\leq K$. Given the trivial sign  and pattern label permutation invariances, we can without loss of generality  consider only non-negative magnetizations, and look for solutions where $m^{\mu}=m$ for $\mu=1\leq K$ and zero otherwise. We then find
\bea\label{m1}
m&=& \sum_{k=-\infty}^{\infty}\pi(k)\int \!{\rm D}z ~\tanh[\beta c(m+m k+z\sqrt{\a r})]
\eea
 with $\pi(k)$ given in (\ref{eq:pik}).  We can now use the manipulations employed in the previous section, to find
\bea
m&=&\left\langle \frac{k}{\phi} \int\!{\rm D}z~ \tanh[\beta c(m k+z\sqrt{\a r})]\right\rangle_k
\label{altern}\\
q&=&\left\langle  \int\!{\rm D}z ~\tanh^2[\beta c(m k+z\sqrt{\a r})]\right\rangle_k,\label{qeq}\\
r&=&\frac{q}{[1-\beta c(1-q)]^2}.\nn	
\eea
 The corresponding free energy assumes the form
\bea
\beta\hat{f}_{\rm RS}(m,q,r)&=& -\log 2
+ \frac{1}{2}\a r(\b c)^2 (1\!-\!q)
+\frac{1}{2}\beta c^2\phi m^2\!
- \frac{\alpha}{2}\Big(\frac{\beta c q}{1\!-\!\beta c(1\!-\!q)}
\!-\!\log[1\!-\!\beta c(1\!-\!q)]\Big)
\nonumber
\\&&
- \Big\langle \int\!{\rm D}z ~\log
\cosh[ \b c(mk
+z\sqrt{\alpha r})]
\Big\rangle_{k}
\eea
Note that we recover the equations of the medium storage regime simply by putting $\a=0$.

\subsection{The zero noise limit}

\begin{figure}
\vspace*{-3mm}
\centering
\includegraphics[width=130mm]{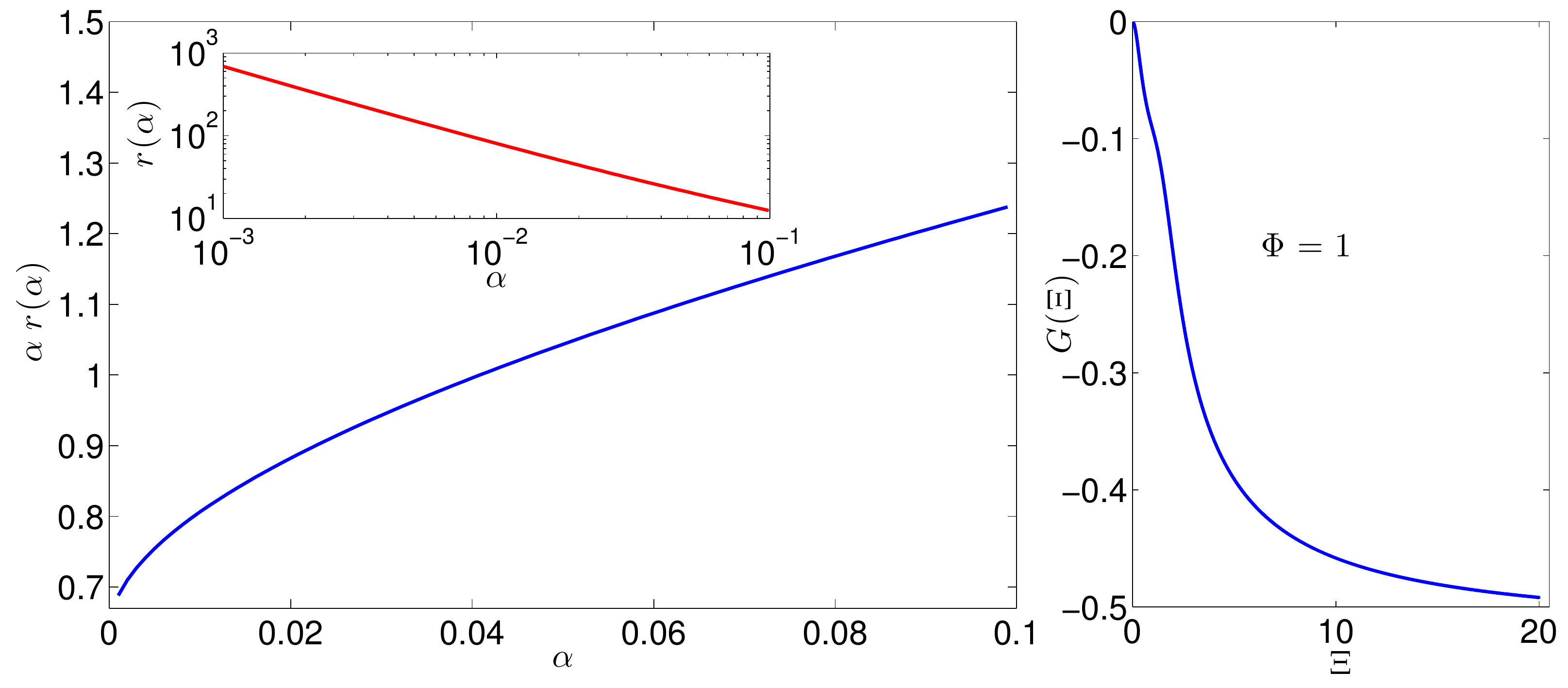}
\caption{Left panel: Behavior of $\alpha r(\alpha)$ versus $\alpha$ in the spin-glass state (the inset shows only $r(\alpha)$ versus $\alpha$), as calculated from the RS order parameter equations. This shows that $r(\a)$ goes to infinity as $\a$ approaches zero, such that $\a r(\a)$ remains positive; this means that the noise due to non-condensed patterns can never be neglected. Right panel: behavior of the function $G(\Xi)$ versus $\Xi$. Since $G(\Xi)<0$ for $\alpha>0$, equation (\ref{eq:ximap}) cannot have a solution for $\alpha>0$, and hence no pattern recall is possible even at zero noise.}
\label{fig:Gxi}
\end{figure}

We now show that in the high storage case the system behaves as a spin glass, even in the zero temperature limit $\beta\to\infty$  where the retrieval capability should be largest.   From $(\ref{qeq})$ we deduce that $q\to 1$ in the zero noise limit, while the quantity $C=\beta c(1-q)$ remains finite. Let us first send $\beta\to\infty$ in  equation (\ref{altern}):
\bea\label{mt0}
m&=&\Big\langle \frac{k}{\phi} \int\!{\rm D}z~ {\rm sgn}\Big[m k+\frac{z\sqrt{\a}}{1\!-\!C}\Big]\Big\rangle_{k}
=~
\Big\langle  \frac{k}{\phi}{\rm Erf}\Big(\frac{mk(1\!-\!C)}{\sqrt{2\a}}\Big) \Big\rangle_{k},
\label{eq:mzeroT}
\eea
with the error integral ${\rm Erf}(x)=(2/\sqrt{\pi})\int_0^x \rmd t~ \rme^{-t^2}\!\!$. A second equation for the pair $(m,C)$ follows from (\ref{qeq}):
\bea
C&=&\lim_{\beta\to\infty}\beta c\Big\langle 1-\int\!{\rm D}z~ \tanh^2[\beta c(m k+z\sqrt{\a r})] \Big\rangle_{k}\nn\\
&=&\lim_{\beta\to\infty}\frac{\partial}{\partial m}\Big\langle \frac{1}{k}\int\!{\rm D}z ~\tanh\Big[\beta c\Big(m k+\frac{z\sqrt{\a q}}{1\!-\!C}\Big)\Big] \Big\rangle_{k},\nn\\
&=& \frac{\partial}{\partial m}\left\langle \frac 1 k {\rm Erf}\Big(\frac{mk(1\!-\!C)}{\sqrt{2\a}}\Big)\right\rangle_{k}
\nn\\
&=&\sqrt{\frac{2}{\a\pi}}(1\!-\!C)~\Big\langle \!\exp\Big(\!-\!\frac{m^2k^2(1\!-\!C)^2}{2\a}\Big) \Big\rangle_{k}
\label{eq:CzeroT}
\end{eqnarray}
We thus have two coupled nonlinear equations (\ref{eq:mzeroT},\ref{eq:CzeroT}), for the two zero temperature order parameters $m$ and $C$. They can be further reduced
by introducing the variable $\Xi=m(1\!-\!C)/\sqrt{2\alpha}$, with which we obtain
\be
m=\left\langle \frac k {\phi} {\rm Erf}(k \Xi)\right\rangle_k
\ee
and rewriting $\Xi=m(1\!-\!C)/\sqrt{2\alpha}$ gives
\be
C=1-\frac{\sqrt{2\a} \Xi}{m}=1-\sqrt{2\a} \Xi~\Big\langle \frac k {\phi} {\rm Erf}(k \Xi)\Big\rangle_k^{-1}.
\ee
Using (\ref{eq:CzeroT}) and excluding the trivial solution $\Xi=0$ (which always exists, but represents the spin glass state without pattern recall) we obtain after some simple algebra just a single equation, to be solved for $\Xi$:
\be
\sqrt{2\a}=G(\Xi)=
\frac{1}{\Xi}\Big\langle \frac k {\phi} {\rm Erf}(k \Xi)\Big\rangle_k-\frac{2}{\sqrt{\pi}}\Big\langle \rme^{-k^2\Xi^2}\Big\rangle_k
\label{eq:ximap}
\ee
One easily shows that
\bea
&&
\lim_{\Xi\to 0}G(\Xi)=0,~~~~~~
\lim_{\Xi\to\infty}G(\Xi)= -\frac{2}{\sqrt{\pi}}\pi(0|\phi).
\eea
In fact further analytical and  numerical investigation reveals that for $\Xi>0$ the function $G(\Xi)$ is strictly negative; see Figure \ref{fig:Gxi}.  Hence there can be no $m\neq 0$ solution for $\alpha>0$,  so the system cannot recall the patterns in the present scaling regime $N_B=\alpha N_T$.

\section{Conclusions}

The immune system is a marvellous complex biological entity,  able to execute reliably a number of very difficult tasks that allow  living beings to survive in competitive interaction with a living environment. To accomplish this it relies on a huge ensemble of functions and agents. In particular, the adaptive part of the immune system relies on a broad ensemble of cells, e.g. B and T lymphocytes, and of chemical messengers, e.g. antibodies and cytokines.
As for lymphocytes, one can distinguish between an `effector branch', consisting of B-cells and killer T-cells, and an `organizational branch', which coordinates the  operation of the effector branch and consists mainly  of helper and regulator T-cells. The latter control the activity of the effector branch through a rich and continuous exchange of cytokines, which are specific chemical messengers which elicit or suppress effector actions.

From a theoretical point of view, a fascinating ability of the immune system is its simultaneous management, by helpers and suppressors, of several B-clones at once; this is a key ability, as it implies the ability to defend the host from simultaneous attacks by several pathogens.
Indeed, we investigated this ability in the present study, as an emergent, collective, feature of a spin glass model of the immune network, that describes the adaptive response performed by B-cells under the coordination of helpers and suppressors. In particular, the focus of this paper is on the ability of the T-cells to coordinate an extensive number of B-soldiers, by fine-tuning the load of clones and the degree of dilution in the network. However, it is worth considering this parallel processing capability also from a slightly different perspective. Beyond the interest in multiple clonal expansions (which, in our language, is achieved trough signalling by $+1$ cytokines), the quiescence signals that are sent to the B-clones that are not expanding (which, in our language, is achieved trough signalling by $-1$ cytokines) is fundamental for homeostasis. In fact, B-cells that are not receiving a significant amount of signals undergo a depauperation process called ''anergy"  \cite{goodnow1}\cite{goodnow2} and eventually die. Hence, in the present multitasking network, the capability of signaling simultaneously to all clones is fundamental, and with implications beyond solely the management of simultaneous clonal expansions;  we emphasize that within our approach  this is achieved in a rather natural way.

We first assumed that the number $N_B$ of B-cells scales with the number $N_T$ of T-cells as $N_B = \alpha N_T^{\delta}$, with $\delta < 1$,  and we modeled the interaction between B cells and T cells by means of an extremely diluted bipartite spin-glass where the former are addressed only by a subset of T cells whose cardinality scales like $N_T^{1-\gamma}$, with $\gamma \leq 1$.
We proved that this system is thermodynamically equivalent to a diluted monopartite graph, whose topological properties are shown to depend crucially on the parameters $\gamma$ and $\delta$. In particular, when $\gamma \geq \delta$ the graph is fragmented into multiple disconnected components, each forming a clique or a collection of cliques typically connected via a bridge. Each clique corresponds to a pattern and this kind of arrangement easily allows for the simultaneous recall of multiple patterns. On the other hand, when $\gamma < \delta$, the effective network can exhibit a giant component, which prevents the system from simultaneous pattern recall.
These results on the topology of the immune network are then approached from a statistical mechanics angle:
we analyse the operation of  the system as an effective equilibrated stochastic process of interacting helper cells.
We find that for $\g>\d$ the network is able to  retrieve perfectly all the stored patterns simultaneously, in perfect agreement with the topology-based prediction.
When the load increases, i.e. when $N_B$ becomes larger (so the exponent $\delta$ is increased), overlaps among bit entries of the `cytokine patterns' to be recalled become more and more frequent, and this gives rise to a new source of non-Gaussian interference noise that is non-negligible for $\g\leq\d$.  If $\g=\d$ the system is still able to retrieve all the patterns, but with a decreasing recall overlap.
In the high storage case,  for $\d=1$, the network starts to feel also the Gaussian noise due to non-condensed patterns, and this is found to destroy the retrieval states. Here  the system behaves as a spin-glass, from which we deduce that an extremely diluted B-H network (i.e. one with $\g<1$) is insufficiently diluted to sustain a high pattern load.
Our predictions and results are tested against numerical simulations  wherever possible, and we consistently find perfect agreement.

Despite the fact that it is experimentally well established that helpers are much more numerous than B-cells, their relative sizes are still comparable in a statistical mechanical sense. The biological interest lies in the high storage regime, where the maximum number of pathogens can be fought simultaneously.  From the present study we now know that to bypass the spin-glass structure of the phase space at this load level, a projection of the model into a finite-connectivity topology ($\g=1$) is required. This, remarkably, is also in agreement with the biological picture of highly-selective  touch-interactions among B and T cells. It is both welcome and encouraging that both biological data and statistical mechanical theory have now converged to the same suggestion: that the most efficient and biologically most plausible operation regime is likely to be that of finite connectivity for the effective helper-helper immune network. This must therefore be the direction of the next stage of our research programme.

\section*{Acknowledgements}

EA, AB and DT acknowledge the FIRB grant RBFR08EKEV and Sapienza Universit\'{a} di Roma for financial support.
ACCC is grateful for support from the Biotechnology and Biological Sciences Research Council (BBSRC) of the United Kingdom. DT would like to thank King's College London for hospitality.

\appendix
\section{Topological properties}

\subsection{Rigorous calculation of link probability}\label{sec:topo2}

We consider the bipartite graph $\mathcal{B}$, and denote with $\rho_i$ the number of links stemming from node $i \in V_T$. Note that $\rho_i$ also gives the number of non-null entries in the string $(\xi_i^{1},\ldots,\xi_i^{N_B})$ processed at node $i$, that is
\be
\rho_i = \sum_{\mu=1}^{N_B} |\xi_i^{\mu}|.
\ee
All  entries $\xi_i^\mu$ are i.i.d. variables (\ref{eq:xi}), so for each node the number $\rho_i$ is distributed according to
\begin{eqnarray}
P(\rho|N_B,N_T,\gamma,c) = \Big(\!\!\begin{array}{c}N_B\\\rho\end{array} \!\!\Big)\Big(\frac{c}{2N_T^{\gamma}}\Big)^{\rho} \Big(1 \!-\! \frac{c}{2N_T^{\gamma}} \Big)^{N_B-\rho}
\end{eqnarray}
When considering two distinct nodes $i, j \in V_T$, the number $\ell$ of shared nearest-neighbors corresponds to the number of non-null matchings between the related strings, and this is distributed according to
\be
P(\ell | \rho_i, \rho_j, N_B) = \frac{N_B!}{(N_B+ \ell - \rho_i - \rho_j)! (\rho_i - \ell)! ( \rho_j - \ell)! \ell!} \left[  \left(\!\!\begin{array}{c}N_B\\\rho_i\end{array}\!\!\right)
\left(\!\!\begin{array}{c}N_B\\\rho_j\end{array}\!\!\right)   \right]^{-1}.
\ee
The average $\langle \ell \rangle_{\rho_i,\rho_j}$ then follows as
\be
\langle \ell \rangle_{\rho_i,\rho_j} = \rho_i \rho_j/N_B.
\ee
By further averaging over $P(\rho|N_B,N_T,c,\gamma)$ we get
\be \label{eq:avell}
\langle \ell \rangle = \langle \rho \rangle^2/N_B.
\ee
Fluctuations scale as $\langle \ell^2 \rangle - \langle \ell \rangle^2 \sim \langle \ell \rangle^2$,
where, from the distribution above, $\langle \rho \rangle= c N_B / 2 N_T^{\gamma}$. Upon choosing  $N_B=\alpha {N_T}^{\d}$ we then get $\langle \ell \rangle \sim N_T^{\d -2 \gamma}$, which vanishes if $2 \gamma > \d$.  Two strings of any two nodes apparently do not display significant matching, so there is no link between them, consistent with the results of Section \ref{sec:topo}.

On the other hand, if $2\gamma<\delta$ so that $\langle \ell \rangle \gg 1$, we can approximate $P(J| N_B, N_T, \gamma, c)$ (the probability of two randonly drawn nodes in the effective $N_T$-node graph having a link $J$) with $P(J | \langle \ell \rangle , N_T, \gamma, c)$: the probability that a random walk of length $N_B$  with a waiting probability $p_w$ ends at distance $J$ from the origin is approximated by the probability that a simple random walk of length $\langle \ell \rangle$ ends at the same distance (with proper normalization to account for parity features).
In particular,
\be
P(J=0 | N_B, N_T, \gamma, c) \approx  \Big(\!\begin{array}{c}\langle \ell \rangle\\
\langle \ell \rangle/2\end{array}\!\Big)  2^{-\langle \ell \rangle} \approx \sqrt{2/\pi \langle \ell\rangle}
\ee
(using Stirling's formula in the last step). The expected link probability between two nodes follows as
\be
P(J \neq 0 | N_B, N_T, \gamma, c) = 1 - P(J=0 | N_B, N_T, \gamma, c) \approx 1 - \sqrt{\frac{2}{\pi \alpha}} \frac{N_T^{\gamma - \d/2}}{c},
\ee
 It is easy to see that when $2 \gamma = \d$ the link probability is finite and smaller than $1$, while when $2 \gamma < \d$ it converges to $1$ in the thermodynamic limit, consistent with the results of Section 3.

\subsection{Generating function approach to percolation in the bipartite graph} \label{genfunc}

Let us consider a bipartite graph $\mathcal{B}$, made of two sets of nodes $V_T$ (of size $N_T$) and $V_B$ (of size $N_B$), with both sizes diverging.
The degree distribution for the two parts are $p_k$ and $q_k$, respectively, with $\sum_k p_k k=\mu$ and $\sum_k q_kk=\nu$. Following \cite{newman}, we introduce the following generating functions
\begin{eqnarray}
& f_0(x)= \sum_{k=0}^{N_T} p_k x^k,~~~~&  g_0(x) = \sum_{k=0}^{N_B} q_k x^k,\\
& f_1(x)= \frac{1}{\mu}\frac{\rmd}{\rmd x} f_0(x), ~~~~&
g_1(x) = \frac{1}{\nu} \frac{\rmd}{\rmd x}  g_0(x),
\end{eqnarray}
We note that $f_1(x)$ and $g_1(x)$ are the generating functions for the degree distribution of a vertex reached following a randomly chosen edge (here the degree does not include the link along which we arrived). One always has
 $\mu / N_T = \nu /N_B$,  and  $f_0(1) = g_0(1) = f_1(1) = g_1(1) = 1$ (by construction).

Next we introduce dilution. We define the matrix $\textbf{t}$, whose element $t_{k\ell}$ represents the probability that a directed link going from a node in part $k$  to a node in part $\ell$ exists. For bipartite graphs, $\textbf{t}$ is simply a $2 \times 2$ matrix with zero  diagonal entries.
We can now write the generating functions for the distributions of occupied edges attached to a vertex chosen randomly  as follows \cite{newman}:
\begin{eqnarray}
f_0(x | \textbf{t}) = f_0(1+(x-1) t_{12}),~~~~~~
f_1(x | \textbf{t}) = f_1(1+(x-1) t_{12}),
\\
g_0(x | \textbf{t}) = g_0(1+(x-1) t_{21} ),~~~~~~
g_1(x | \textbf{t}) = g_1(1+(x-1) t_{21}).
\end{eqnarray}
Let us now consider a node $i \in V_T$,  with $z_i$ neighbors (where $z_i$ is distributed according to $p_k$). Due to the dilution, only a fraction of the links that could connect to $i$ will be present. The nodes in the second part that are reached from $i$ will, in turn, have a number of links hitting some nodes in $V_T$. The generating function $F_0(x | \textbf{t})$ of the distribution of nodes in the first part which are involved in both steps is
\begin{eqnarray}
F_0(x | \textbf{t} ) &=& \sum_{m=0}^{\infty}   \sum_{k=m}^{\infty} p_k \Big(\!\!\begin{array}{c}k\\[-0.5mm]m\end{array}\!\!\Big) t_{12}^m (1- t_{12})^{k-m} [g_1(x;  \textbf{t})]^m \nonumber
\\
&=& f_0(g_1(x| \textbf{t}) |\textbf{t}) = f_0(1 + ( g_1(x| \textbf{t})- 1) t_{12}).
\label{eq:F0}
\end{eqnarray}
In fact, in the expansion of $[g_1(x;  \textbf{t})]^m$, the coefficient of $x^n$ is simply the probability that $m$ randomly reached nodes are connected to a set of $n$ other nodes. If we choose an edge rather than a node we have, analogously
\be \label{eq:F1}
F_1(x | \textbf{t} ) =  f_1(1 + ( g_1(x| \textbf{t})- 1) t_{12}),
\ee
The corresponding generating functions found upon starting with a note in the part $V_B$ have analogous definitions, and will be written as  $G_0$ and $G_1$.

The generating function $H_0$ for the distribution $P(s| \textbf{t} )$ of the size $s$ of the components (connected sub-graphs) which one can detect is $H_0(x | \textbf{t}) = \sum_s P(s | \textbf{t}) x^s$. Similarly,  $H_1(x | \textbf{t})$ will be the generating function for the size of the cluster of connected vertices that we reach by following a randomly chosen vertex.  We note that  in the highly-diluted regimes we can exploit the fact that the probability of finding closed loops is $O(N_T^{-1})$ (so $H_0$ and $H_1$ do not include the giant component), which allows us to write the explicit expressions
\begin{eqnarray}
H_0(x | \textbf{t}) =x F_0(H_1(x|\textbf{t})|\textbf{t}),~~~~~~
H_1(x|\textbf{t}) = x F_1(H_1(x|\textbf{t})|\textbf{t}).
\end{eqnarray}
This then gives for the average cluster size:
\be
\langle s \rangle = H_0'(1|\textbf{t}) = 1 + F_0'(1|\textbf{t}) H_1'(1|\textbf{t}) = 1 - \frac{F_0'(1|\textbf{t})}{1-F_1'(1|\textbf{t})},
\ee
where we used $H_1'(1|\textbf{t})= 1 + F_1'(1 |\textbf{t}) \, H_1'(1|\textbf{t})$.
As for $F_0$ and $F_1$, recalling Eqs.~(\ref{eq:F0}) and (\ref{eq:F1}) we get
$F_0'(1|\textbf{t}) = f_0(g_1(1|t_{21})|t_{12}) g_1'(1|t_{21}) = f_0'(1|t_{12}) g_0'(1)t_{21}$, and analogous formulae for $F_1(1|\textbf{t})$. Therefore,
\be
\langle s \rangle = 1 + \frac{t_{12}t_{21} f_0'(1) g_0'(1)}{1 - t_{12}t_{21} f_1'(1)g_1'(1)}.
\ee
This expression diverges for
\be
t_{12}t_{21} = \frac{1}{f_1'(1) g_1'(1)} = \frac{\mu \nu}{\big(\sum_j j(j\!-\!1)p_j \big)\big(\sum_k k(k\!-\!1)q_k\big)}.
\ee
signalling the phase transition at which a giant-component first appears.
Assuming $\textbf{t}$ to be symmetric, as we are interested in equilibrium statistical-mechanical descriptions of the system, the left hand side of the previous equations simplifies into $t^2 \equiv t_{12}^2$. As for the degree distributions $p_k$ and $q_k$, the case considered in the main text is the easiest one, viz. an originally fully-connected bipartite network that has been progressively diluted, in such a way that $p_k =  \delta_{k,N_T-1}$ and $q_k = \delta_{k,N_B-1}$. Hence, we immediately find
\be
t^2 \approx 1/N_T N_B.
\ee
As mentioned earlier, the distribution for the size of the small components reached from a randomly chosen node $i \in V_T$ has generating function $H_0(x|\textbf{t}) = x F_1(H_1(x|\textbf{t})|\textbf{t})$.

\begin{figure}[t]
\vspace*{-4mm}
 \begin{center}
\includegraphics[width=95mm]{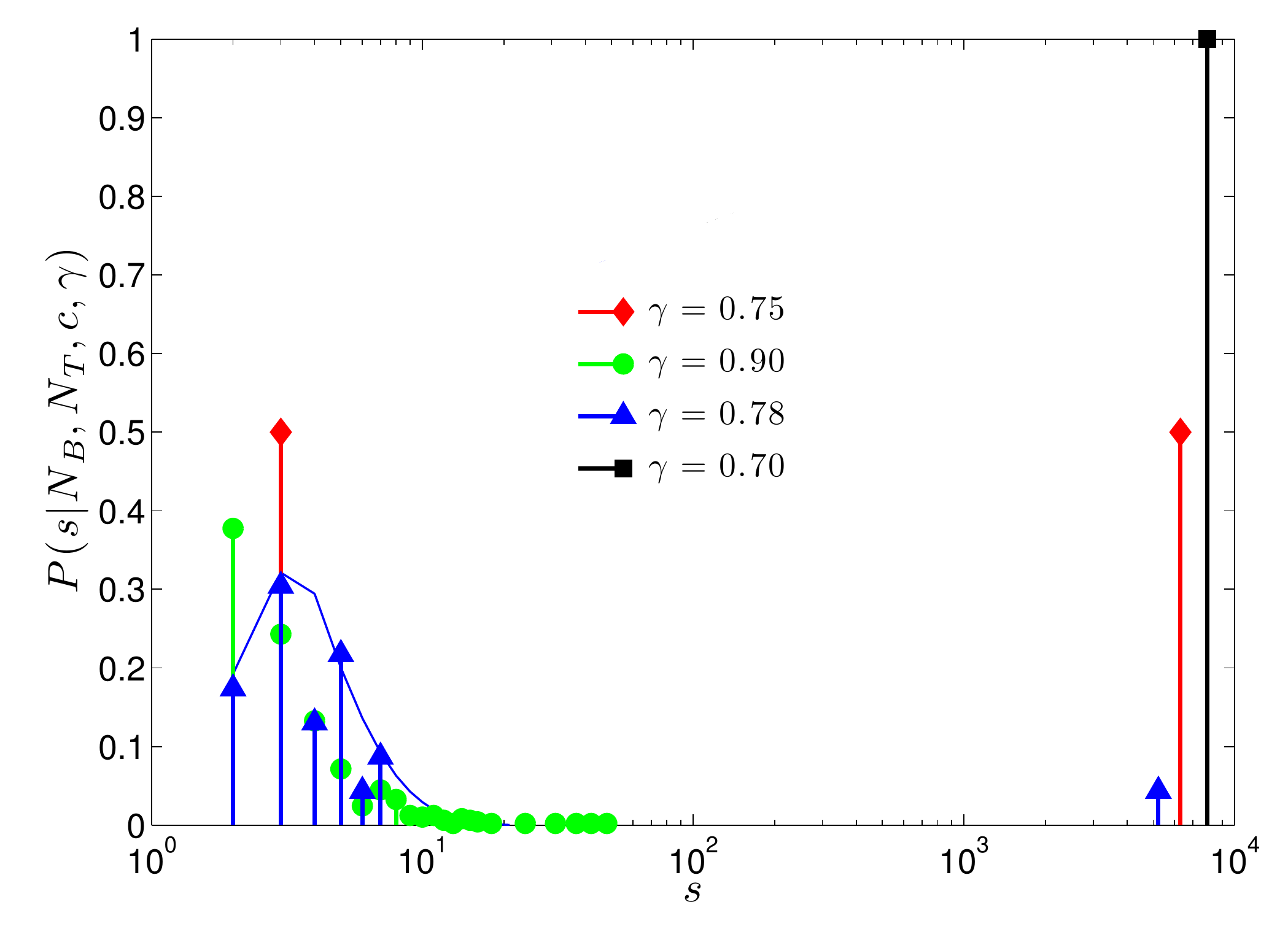}
\caption{\label{fig:Cammello} Distribution for the cluster size $P(s|N_B,N_T,c,\gamma)$ measured numerically over a simulated graph $\mathcal{G}$ with $N_T=10^4$ nodes. The parameters $\d=1$, $\alpha=0.5$ and $c=1$ are kept fixed, while $\gamma$ is varied (see legend). In the interest of  clarity, we plotted the analytical estimate of Eqs.~\ref{eq:PsLong} and \ref{eq:PsLong_approx} only for $\gamma=0.78$.}
\end{center}
\end{figure}

The generating function approach implicitly assumes that small components are always tree-like, i.e. the probability of finding closed loops in finite components is negligible in the large system size limit. Hence, if $t = 1-c/N_T^{\gamma}$, then $\gamma \geq 1$, so we are in effect considering how the system approaches the percolation threshold from the underpercolated regime. Therefore, $H_1(x|\textbf{t}) \approx x$ in such a way that
\be \label{eq:F1approx}
F_1(x|\textbf{t}) = \{ 1 - t + t[1+(x-1)t]^{N_B-2} \}^{N_T-2}.
\ee
Upon substituting (\ref{eq:F1approx}) into the expression for $H_0(x| \textbf{t})$, followed by a numerical inverse Laplace transform, we obtain clear evidence that, when $s$ is small, the leading term for $P(s|\textbf{t})$ decreases exponentially with $s$.  This prediction is confirmed by numerical simulations, see  the right panel of Fig 4.
More generally, the distribution of component sizes can be written as
\be \label{eq:PsLong}
P(s|N_B,N_T,c,\gamma) = \Big(\!\!\begin{array}{c}N_T\\s\end{array}\!\!\Big) \sum_{l=1}^{N_B}
\Big(\!\!\begin{array}{c}N_B\\l\end{array}\!\!\Big)
(1-p)^{s(N_B-l)} (1-p)^{l(N_T-s)} C(l,s),
\ee
where we accounted for the probability of choosing a component ($\subseteq V_T$) of size $s$ linked through, overall, $l$ nodes ($\subseteq V_B$), and for the probability that this sub-graph is disconnected from the remaining nodes; $C(l,s)$ is the probability that a subgraph made of $s+l$ elements is connected.
Now, the $s$ nodes $\in V_T$ can be partitioned into sub-graphs $\{ s_1, s_2, ..., s_{l}\}$, where $s_k$ includes all nodes linked to exactly $k$ nodes among the $l$ selected.
Therefore, we can write
\be
C(l,s) =\sum_{s_1,\ldots, s_l} \frac{s! ~\delta_{s,\sum_k |s_k|}}{\prod_{k=1}^{l} |s_k|!} \prod_{k=1}^l [p^k(1-p)^{l-k}]^{|s_{k}|} 
\ee
A simple upper bound for $C(l,s)$  follows by imposing that all $s+l$ nodes are connected to at least one node
\be  \label{eq:PsLong_approx}
C(l,s) \leq \tilde{C} (l,s) = [1 - (1-p)^l]^s [1 - (1-p)^s]^l,
\ee
This does not imply that the whole sub-graph is connected, but the bound is a good approximation  when the link probability is either low or high.
Using this approximation and the expression $p=c/N_T^{\gamma}$, we find that: for relatively small $\gamma$ only the case $s \sim O(N_T)$ and $l \sim O(N_B)$ has non-vanishing probability, for relatively large $\gamma$ only the case with $s$ and $l$ finite has non-vanishing probability, and for intermediate values of $\gamma$ both these  extreme cases are possible.
In Fig.~\ref{fig:Cammello} we show a comparison between analytical and numerical results.

\section{Appendix B: free energy evaluation using the replica method}

In this appendix we  calculate  the free energy  per spin of the system characterised by the Hamiltonian  (\ref{eq:effHamiltonian}),
within the replica-symmetric (RS) ansatz, for the scaling regime $N_B=\alpha N_T$.
 Let us start by introducing the partition function $Z_{N_T}(\beta,\xi)$ and the disorder-averaged free energy $\overline{f}$:
\begin{eqnarray}
Z_{N_T}(\beta,\xi)&=&\sum_{\bsigma}\rme^{\frac{1}{2}\b N_T^{-\tau}\sum_{i,j=1}^{N_T}\sum_{\mu=1}^{N_B}\xi^{\mu}_i\xi^{\mu}_j\s_i\s_j}
\\
\overline{f}&=&-\lim_{N_T\to\infty}\frac 1 {\b N_T}\overline{\log Z_{N_T}(\beta,\xi)},
\end{eqnarray}
where $\overline{\cdots}$ denotes averaging over the randomly generated $\{\xi_i^\mu\}$.
If we use the replica identity $\overline{\log Z}=\lim_{n\to 0}n^{-1}\log \overline{Z^n}$, and separate the contributions from the $K$ condensed patterns from those of  the $\alpha N_T\!-\!K$ non-condensed ones we get
\bea\label{repl}
\overline{f}&=&-\lim_{N_T\to\infty}\lim_{n\to 0} \frac{1}{\b n N_T} \log \sum_{\bsigma^1,\cdots,\bsigma^n}\overline{\rme^{\frac{1}{2}\b N_T^{-\tau}\sum_{i,j=1}^{N_T}\sum_{\mu=1}^{N_B}\sum_{\a=1}^n \xi^{\mu}_i\xi^{\mu}_j \s^{\a}_i\s^{\a}_j}}\nn\\
&=&-\frac{1}{\b}\log 2 -\lim_{N_T\to\infty}\lim_{n\to 0} \frac{1}{\b n N_T} \log
\Big\langle \rme^{\frac{1}{2}\b N_T^{-\tau}\sum_{\mu=1}^K\sum_{\a=1}^n (\sum_{i=1}^{N_T}\xi^{\mu}_i \s^{\a}_i)^2}
\nonumber
\\
&&\hspace*{50mm}\times
\overline{\rme^{\frac{1}{2}\b N_T^{-\tau}\sum_{\mu>K}^{N_B}\sum_{\a=1}^n (\sum_{i=1}^{N_T}\xi^{\mu}_i \s^{\a}_i)^2}}\Big\rangle_{\bsigma^1,\ldots,\bsigma^n}.
\eea
We compute the non-condensed contributions first, using the standard tool of Gaussian linearisation, and the usual short-hands ${\rm D}z=(2\pi)^{-1/2}\rme^{-z^2/2}\rmd z$ and ${\rm D}\bz=\prod_{\alpha=1}^n{\rm D}z_\alpha$:
\begin{eqnarray}
\Xi~&=& ~
\overline{\rme^{\frac{1}{2}\b N_T^{-\tau}\sum_{\mu>K}\sum_{\a=1}^n (\sum_{i=1}^{N_T}\xi^{\mu}_i \s^{\a}_i)^2}}
=
\left[\overline{\rme^{\frac{1}{2} \b N_T^{-\tau}\sum_{\a=1}^n (\sum_{i=1}^{N_T}\xi_i \s^{\a}_i)^2}}\right]^{N_B-K}
\nn
\\
&=&\left[\int\! {\rm D}\bz ~\overline{\rme^{\sqrt{\b}N_T^{-\tau/2}\sum_{\a=1}^n z_\alpha\sum_{i=1}^{N_T}\xi_i \s^{\a}_i}}\right]^{N_B-K}
\nn
\\
&=&\left[\int\!{\rm D}\bz ~\prod_{i=1}^{N_T}\Big(
1\!-\!cN_T^{-\g}\!+\!cN_T^{-\g}\cosh(\sqrt{\b}N_T^{-\tau/2}\sum_{\a=1}^n\s^{\a}_i z_{\a})\Big)\right]^{N_B-K}
\nn
\\
&=&\left[\int\! {\rm D}\bz~ \prod_{i=1}^{N_T}\Big[
1\!+\!\frac{1}{2}\beta cN_T^{-\g-\tau}\big(\sum_{\a=1}^n\s^{\a}_i z_{\a}\big)^2+\order(N_T^{-2\tau-\gamma})
\Big]
\right]^{N_B-K}
\nn
\\
&=&\left[\int\! {\rm D}\bz ~
\rme^{\frac{1}{2}\beta cN_T^{-\g-\tau}\sum_{\a,\beta=1}^n z_\alpha z_\beta\sum_{i=1}^{N_T}\s^{\a}_i \s^\beta_i
+\order(N_T^{1-2\tau-\gamma})}
\right]^{N_B-K}.
\eea
Now it is evident, as in our earlier calculations,  that the correct scaling for large $N_T$ requires choosing $\tau=1-\g$.
For the correction term in the exponent this gives $\order(N_T^{1-2\tau-\gamma})=\order(N_T^{\gamma-1})$, which is indeed vanishing since $\gamma<1$.
We now arrive at
\begin{eqnarray}
\Xi~&=&~\exp\Big\{(N_B\!-\!K)\log \int\! {\rm D}\bz ~
\rme^{\frac{1}{2}\beta cN_T^{-1}\sum_{\a,\beta=1}^n z_\alpha z_\beta\sum_{i=1}^{N_T}\s^{\a}_i \s^\beta_i
+\order(N_T^{\gamma-1})}
\Big\}
\label{eq:Xi}
\end{eqnarray}
We next introduce $n^2$ parameters $\{q_{\a\b}\}$ and their conjugates $\{\hat{q}_{\a\b}\}$, by  inserting partitions of unity:
\be\label{partun}
\hspace*{-5mm}
1=\prod_{\a\b}\int\! \rmd q_{\a\b}~\d\big(q_{\a\b}-\frac{1}{N_T} \sum_{i=1}^{N_T}\s^{\a}_i\s^{\b}_i\big)
=\int\!\!\Big[\prod_{\a\b} \!\frac{\rmd q_{\a\b}\rmd\hat{q}_{\a\b}}{2\pi/N_T}\Big]
\rme^{\rmi N_T\sum_{\a,\b}\hat{q}_{\a\b}(q_{\a\b}-\frac{1}{N_T} \sum_{i}\s^{\a}_i\s^{\b}_i)}.
\ee
Substituting (\ref{partun}) into (\ref{eq:Xi}) gives the contribution to the partition function of non-condensed patterns:
\begin{eqnarray}
\Xi&=&
\int\!\!\Big[\prod_{\a\b} \!\rmd q_{\a\b}\rmd\hat{q}_{\a\b}\Big]
\rme^{\rmi N_T\sum_{\a,\b}\hat{q}_{\a\b}q_{\a\b}+(N_B\!-\!K)\log \int\! {\rm D}\bz ~
\rme^{\frac{1}{2}\beta c\sum_{\a,\beta=1}^n z_\alpha q_{\alpha\beta}z_\beta
}+\order(N_T^{\gamma})}
\nonumber
\\[-2mm]
&&\hspace*{70mm}
\times
\rme^{-\rmi \sum_{i}\sum_{\a,\b}\s^{\a}_i\hat{q}_{\a\b}\s^{\b}_i}.
\label{eq:guzai}
\end{eqnarray}
The contribution from condensed pattern, see (\ref{repl}), is
\be
\rme^{\frac{1}{2}\b N_T^{\gamma-1}\sum_{\mu\leq K}\sum_{\a=1}^n (\sum_{i=1}^{N_T}\xi^{\mu}_i \s^{\a}_i)^2}=
\int\! {\rm D}\bm~  \rme^{\sqrt{\b}N_T^{(\g-1)/2}\sum_{\mu\leq K}\sum_{\a=1}^n \sum_{i=1}^{N_T}\xi^{\mu}_i \s^{\a}_im^{\mu}_{\a}},
\ee
with $\bm=\{m_\alpha^\mu\}\in\R^{nK}$.
If we rescale $m^{\mu}_{\a}\to c\sqrt{\b}N_T^{(1-\g)/2} m^{\mu}_{\a}$ this becomes
\be\label{partdu}
\Big(c^2\b N_T^{1-\g}\Big)^{\!nK/2}\!
\int\! \rmd\bm~ \rme^{
-\frac{1}{2}\beta c^2 N_T^{1-\g} \bm^2
 +\b c\sum_{\mu\leq K}\sum_{\a=1}^n \sum_{i=1}^{N_T}\xi^{\mu}_i \s^{\a}_im^{\mu}_{\a}}.
\ee
Inserting (\ref{eq:guzai},\ref{partdu}) into (\ref{repl}) gives the following expression for the free energy per spin:
\bea
\overline{f}&=&-\frac{1}{\b}\log 2
 -\lim_{N_T\to\infty}\Big\{\frac{K/2}{\b  N_T} \log(c^2\b N_T^{1-\g})
\nonumber
\\
&&
+\lim_{n\to 0} \frac{1}{\b n N_T} \log
\int\! \rmd\bm\Big[\prod_{\a\b} \!\rmd q_{\a\b}\rmd\hat{q}_{\a\b}\Big]
\nonumber
\\
&&\times
\rme^{N_T\Big[\rmi \sum_{\a,\b}\hat{q}_{\a\b}q_{\a\b}+\frac{N_B\!-\!K}{N_T}\log \int\! {\rm D}\bz ~
\rme^{\frac{1}{2}\beta c\sum_{\a,\beta=1}^n z_\alpha q_{\alpha\beta}z_\beta
}
-\frac{1}{2}\beta c^2 N_T^{-\g} \bm^2\Big]}
\nonumber
\\[-0mm]
&&\hspace*{0mm}
\times
\prod_{i=1}^{N_T}
\Big\langle
\rme^{
 \b c\sum_{\mu\leq K}\sum_{\a=1}^n \xi^{\mu}_i \s^{\a}_im^{\mu}_{\a}
-\rmi \sum_{\a,\b}\s^{\a}_i\hat{q}_{\a\b}\s^{\b}_i}.
\Big\rangle_{\sigma_i^1,\ldots,\sigma_i^n}
\Big\}
\eea
The number of order parameters being integrated over is of order $K$, so corrections to the saddle-point contribution will be of order
$\order(K\log N/N)$.  To proceed via steepest descent we must therefore impose $K\ll N_T/\log N_T$. Since also the energy term $N_T^{-\g} \sum_{\mu\leq K}\bm^2$ should be of order one, as well as the individual components of $\bm$, the only natural choice is
$K=\order(N^\gamma)$.
Under this scaling condition we then find
\bea
\overline{f}&=&-\frac{1}{\b}\log 2
 -\lim_{K\to\infty}
\lim_{n\to 0} \frac{1}{\b n }
{\rm extr}_{\bm,q,\hat{q}}\hat{f}(\bm,\{q,\hat{q}\})
\end{eqnarray}
with
\begin{eqnarray}
\hat{f}(\bm,\{q,\hat{q}\})&=&
\rmi \sum_{\a,\b}\hat{q}_{\a\b}q_{\a\b}+\alpha\log \int\! {\rm D}\bz ~
\rme^{\frac{1}{2}\beta c\sum_{\a,\beta=1}^n z_\alpha q_{\alpha\beta}z_\beta
}
-\frac{\beta c^2}{2N_T^\gamma}\sum_{\alpha=1}^n\sum_{\mu\leq K}(m_\alpha^\mu)^2
\\&&
+\Big\langle\log
\Big\langle
\rme^{
 \b c\sum_{\mu\leq K}\sum_{\a=1}^n \xi^{\mu}\s^{\a}m^{\mu}_{\a}
-\rmi \sum_{\a,\b}\s^{\a}\hat{q}_{\a\b}\s^{\b}}.
\Big\rangle_{\sigma^1,\ldots,\sigma^n}\Big\rangle_{\xi}
\eea
Now we can use the replica symmetry ansatz, and demand that the relevant saddle-point is of the form
\be
m^{\mu}_{\a}=m^{\mu},~~~~~ q_{\a\b}=\d_{\a\b}+q(1\!-\!\d_{\a\b}),~~~~~ \hat{q}_{\a\b}=\frac{\rmi\a(\b c)^2}{2}[R \d_{\a\b}+r(1\!-\!\d_{\a\b})],
\ee
From now on we will denote $\bm=(m^1,\ldots,m^K)$ and $\bxi=(\xi^1,\ldots,\xi^K)$. After some simple algebra we can take the limit $n\to 0$, and   find that
our free energy simplifies to
\bea
\beta\overline{f}_{\rm RS}&=&\lim_{N_T\to\infty}
{\rm extr}_{m,q,r}~\hat{f}_{\rm RS}(\bm,q,r)
\end{eqnarray}
with
\begin{eqnarray}
\hat{f}_{\rm RS}(\bm,q,r)&=& -\log 2
+ \frac{1}{2}\a r(\b c)^2 (1\!-\!q)
+\frac{\beta c^2}{2N_T^\gamma}\bm^2
- \frac{\alpha}{2}\Big(\frac{\beta c q}{1\!-\!\beta c(1\!-\!q)}
-\log[1\!-\!\beta c(1\!-\!q)]\Big)
\nonumber
\\&&
- \Big\langle \int\!{\rm D}z ~\log
\cosh[ \b c(\bm\cdot\bxi
+z\sqrt{\alpha r})]
\Big\rangle_{\xi}
\eea

\end{document}